\documentclass[a4paper,11pt]{article}
\pdfoutput=1 % if your are submitting a pdflatex (i.e. if you have
\usepackage{jheppub} % for details on the use of the package, please\input{WNFieldTheory11.tex}

\usepackage[T1]{fontenc} % if needed
\usepackage{fancyhdr}
\usepackage{mathtools}
\usepackage[vcentermath]{youngtab}
\usepackage{multirow}
\usepackage{enumitem}
\usepackage{xfrac}

\title{\boldmath Field Theory of Primaries in $W_N$ Minimal Models}

\author{Antal Jevicki,}
\author{Junggi Yoon}

\affiliation{Department of Physics, Brown University,\\Providence, RI 02912, USA}

\emailAdd{antal\_jevicki@brown.edu}
\emailAdd{jung-gi\_yoon@brown.edu}

\abstract{ For $W_N$ minimal model CFT at Large $N$, we formulate a nonlinear field theory describing interaction of primaries. A classification of single-trace operators is given first based on which a field theory operating in Fock space is built. A hamiltonian is constructed with the property that it reproduces exactly the spectrum of conformal dimensions of the primaries. This field theory is characterized  by an extra dimension and interactions with $G=1/N$ as the coupling constant. It is seen that this nonlinear representation contains structure parallel to the one of Matrix-vector models.}

\begin{document} 

\begin{flushright}
 {\tt BROWN-HET-1639 \quad}
\end{flushright}
\maketitle
\flushbottom

\newpage
\section{Introduction}

Dualities involving Vasilev's Higher Spin Gravity~\cite{vasiliev1987:1,vasiliev1987:2,vasiliev1987:3,vasiliev1999,vasiliev2003} and Large $N$ conformal field theories~\cite{klebanov2002,sezgin2002} have been a topic of active current investigations~\cite{giombi2010,giombi2011:1,giombi2011:2,Das:2003vw,antal2010, antal2011,Giombi:2012ms,Maldacena2011,Gelfond:2013xt}. The lower dimensional AdS$_3$/CFT$_2$ duality based on $W_N$ minimal models \cite{Gaberdiel:2012uj} is characterized by the high $W_\infty$ symmetry group~\cite{henneaux2010,campoleoni2010,gaberdielhartman2011,campoleoni2011} characteristic of Higher Spin Gravity in AdS$_3$. The correspondence is bolstered by comparison of the spectrum and of the partition function (in leading Large $N$)~\cite{gaberdielgopakumar2011,gaberdielsaha2011,gaberdielgopakumarhartman2011,yin2011:1,gaberdielcandu2012,Gaberdiel:2012uj}. It is expected that again $1/N$ plays the role of the Higher Spin Gravity coupling constant $G$ and that the correspondence persists at full nonlinear level.
In vector model field theories it is relatively simple to formulate a $1/N$ expansion based either on  Feynman diagrams or through an effective bi-local collective field theory~\cite{antal2010, antal2011}. It was seen that this field theoretic formulation carries all the features  of  bulk AdS space-time giving  a  comparison of Hilbert spaces between the two sides of the AdS$_4$/CFT$_3$ duality. One can expect that that the field theoretic construction and the derived Hilbert space will play a useful role in non-perturbative studies of the duality.
When it comes to 2 dimensional $ W_N$ conformal field theories one topic which is not that well understood , is the structure and systematics of the $1/N$ expansion. In practice two different limits were introduced : the Large $N$ t'Hooft and the large $c$ semiclassical limit, leading to similar but not identical dualities. An exact map of Hilbert spaces and the formulation of the duality at interacting level is still a goal.
In this work we begin to do that concentrating on the space of the primary operators of $W_N$ conformal field theories. In this we are motivated by the successful understanding of $1/2$ BPS primaries in $\mathcal{N}=4$ Super Yang Mills theory in terms of Fermion droplet theory which was found both in Yang-Mills and Supergravity descriptions \cite{Corley:2001zk,Lin:2004nb,Jevicki:1998rr}. A central role in this and the AdS$_4$/CFT$_3$ bi-local construction is played by the so-called single-trace operators which are the basic building blocks of the construction. In $U(N)$ group theory models  this role is played by characters (of representations).
In a related publication we give a pedagogical description of this construct with a simple CFT extension involving the $S_N$ orbifold theory.

The content of our paper is as follows. In section~\ref{sectionwnprimary}, we first summarize the basic facts about $W_N$ primary states and describe previous work by Chang and Yin \cite{yin2011,yin2012} on studying their single and multi-trace structure.  We then give a general characterization of all single traces and formulate a scheme for general multi-trace generalization. This scheme involves a Fock space  based on single trace primaries and a nonlinear (collective) Hamiltonian. The significance of a nonlinear Hamiltonian is twofold. First, it represents a field theory in one extra dimension, related to the winding number. Second, it generates a complete set of multi-trace primaries as eigenfunctions in its Hilbert space, through degenerate perturbation theory.It is shown to reproduce  the exact conformal dimensions of all the primaries. The $N$-dependence of this (collective) Hamiltonian is through $1/N$ as a coupling constant parameter. Consequently this description provides a framework for a systematic $1/N$ expansion and is also a first step in a direct construction of bulk Higher Spin theory from CFT.

The interacting field theory constructed can be recognized to have the features of  the well studied matrix-vector model. We elaborate on this in section~\ref{connectiontomv} together with the geometric picture (of interactions) in the theory. In the discussion we mention the role of the Large $c$ limit and summarize a number of outstanding issues.

\section  {$W_N$ Minimal Model and its Primaries}\label{sectionwnprimary}

\begin{figure}[tbp]
\centering
\includegraphics[width=5.5cm]{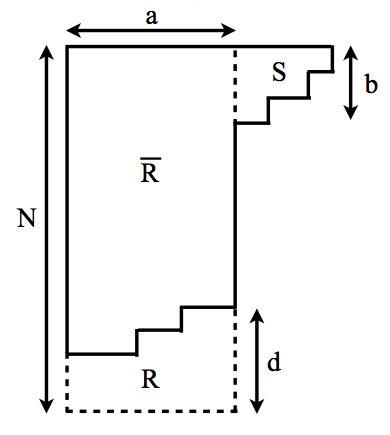}
\caption{\label{youngtableau1} Young tableau $\Lambda=(\overline{R},S)$}
\end{figure}

As is well known, the $W_N$ minimal model can be represented by a diagonal coset WZW model \cite{bouwknegt1993}
\begin{align}
\frac{su(N)_k\oplus su(N)_1}{su(N)_{k+1}}
\end{align}
whose primary states (vertex operators) are  labelled by two Young tableaux of $SU(N)$, $\left(\Lambda_+;\Lambda_-\right)$. The conformal dimension of $\left(\Lambda_+;\Lambda_-\right)$ primary state is exactly given
\begin{align}
&h\left(\Lambda_+;\Lambda_-\right)=\frac{1}{2p(p+1)}\left\{\left|(p+1)\Lambda_+-p\Lambda_-+\rho\right|^2-\left|\rho\right|^2\right\}
\end{align}
where $\rho$ is a Weyl vector and $p=N+k$. We can express the conformal dimension in terms of variables of Young tableau.
\begin{align}
\label{conformaldimension}
\begin{split}
h\left(\Lambda_+;\Lambda_-\right)=&\frac{\lambda}{2}\left(B_+-B_-\right)+\frac{1}{2}\sum_{i=1}^{N-1} \left(r_{i}^+-r_{i}^-\right)^2+\frac{\lambda}{2N}\left(D_+-D_-\right)-\frac{1}{2N}\left(B_+-B_-\right)^2\\
&-\frac{\lambda}{2N^2}\left(B^2_+-B^2_-\right)+\frac{\frac{\lambda^2}{N^2}}{1+\frac{\lambda}{N}}\left(\frac{1}{2}B_-N+\frac{1}{2}D_--\frac{B_-^2}{2N}\right)
\end{split}
\end{align}
where
\begin{align*}
r_i^\pm=&\left(\mbox{The number of boxes in the $i$th row of Young tableau }\Lambda_\pm\right)\\
c_i^\pm=&\left(\mbox{The number of boxes in the $i$th column of Young tableau }\Lambda_\pm\right)\\
B_\pm=&\left(\mbox{The total number of boxes in Young tableau }\Lambda_\pm\right)=\sum_{i=1}^{N-1} r_i^\pm=\sum_{j=1}^\infty c_j^\pm\\
D_\pm=&\sum_{i=1}^{N-1}\left(r_i^\pm\right)^2-\sum_{j=1}^\infty \left(c_j^\pm\right)^2
\end{align*}
and 
\begin{align*}
p\equiv& N+k\;,\qquad\lambda\equiv \frac{N}{N+k}
\end{align*}
The  significance of the above formula is that it  represents an exact result to all orders in $1/N$.
The large $N$ limit  of Gaberdiel and Gopakumar~\cite{gaberdielgopakumarhartman2011} features  representations $\Lambda$ of $SU(N)$ generated by basic fundamental  and anti-fundamental representations. This representation $\Lambda$ can be expressed by two finite Young tableaux, $\overline{R}$ and $S$. (e.g. see figure~\ref{youngtableau1})
\begin{align}
\Lambda_\pm=\left(\overline{R}_\pm,S_\pm\right)
\end{align}
Then, the conformal dimension of $\left(\Lambda_+;\Lambda_-\right)$ can be separated into two conformal dimensions of $\left(R_+;R_-\right)$ and $\left(S_+;S_-\right)$ up to the number of boxes of each Young tableaux. Explicit formulae for the conformal dimension are given in appendix~\ref{conformaldimappendix}. The  $W_N$  CFT primaries can be constructed in the Coulomb gas scheme in terms of free boson vertex operators  like
\begin{align}
\mathcal{O}_{(\Lambda_+;\Lambda_-)}\sim e^{iv\cdot X}\qquad \mbox{where}\quad v=\sqrt{\frac{p+1}{p}}\Lambda_+-\sqrt{\frac{p}{p+1}}\Lambda_-
\end{align}

In the present construction we will deal only with primary operators(states). This will correspond to certain localization of the nonlinear field theory . In $W_N$ minimal models all the remaining states are generated from the primary operators through action of the $W_N$ operators . This will lead to ``dressing'' of the states and the field theory  constructed, which will be addressed in subsequent work.

It will be relevant for our construction to separate the operators into a single-trace and multi-trace ones. This notion is analogous to operators in matrix and vector model field theories,
but it exists in any theory based on invariants under a non-abelian symmetry group. In the WZW model it is not directly visible how this separation is to be performed.  

In \cite{yin2011}, Chang and Yin considered this problem using the different factorization properties and $1/N$ dependence in correlation functions  and have provided examples of the identification in 
low lying cases.  Three point functions in the $W_N$ minimal model, can be in principle calculated through analytic continuation from three point functions of affine Toda theory.

The  operators $\mathcal{O}_{(\Lambda_+;\Lambda_-)}$ are normalized in  two point functions as
\begin{align}
\left<\mathcal{O}_{(\Lambda_+;\Lambda_-)}\left(x\right)\;\mathcal{O}_{(\overline{\Lambda}_+;\overline{\Lambda}_-)}\left(0\right)\right>=\frac{1}{\left|x\right|^{2\Delta}}
\end{align}
where $\Delta$ is the scaling dimension.

The structure of the general  three point functions of the $W_N$ minimal model is
\begin{align}
\left<\mathcal{O}_{(\Lambda_+^1;\Lambda_-^1)}\mathcal{O}_{(\Lambda_+^2;\Lambda_-^2)}\mathcal{O}_{(\Lambda_+^3;\Lambda_-^3)}\right>=\frac{C_{3}\left({(\Lambda_+^1;\Lambda_-^1)},{(\Lambda_+^2;\Lambda_-^2)},{(\Lambda_+^3;\Lambda_-^3)}\right)}{\left|x_{12}\right|^{\Delta_1+\Delta_2-\Delta_3}\left|x_{23}\right|^{\Delta_2+\Delta_3-\Delta_1}\left|x_{31}\right|^{\Delta_3+\Delta_1-\Delta_2}}
\end{align}
where $C_{3}\left({(\Lambda_+^1;\Lambda_-^1)},{(\Lambda_+^2;\Lambda_-^2)},{(\Lambda_+^3;\Lambda_-^3)}\right)$ is a structure constant.
Based on knowledge of these, one can find single-trace operators from three point functions in principle.  For example, suppose that we know two single-trace operators. Then, by calculating a three point function of an unknown operator and the two known single-trace operators, we can find a relation between the unknown operator and the two single-trace operator. 

Starting for example from a trivial identification\footnote{We use different notations of operators from \cite{yin2011} of
\begin{align*}
\psi_0\quad\Longleftrightarrow\quad \phi\;,\qquad {\tiny\yng(1,1)}\quad\Longleftrightarrow\quad \text{A}\;,\qquad\left({\tiny\yng(1)},\overline{{\tiny\yng(1)}}\right)\quad\Longleftrightarrow\quad \text{adj}
\end{align*}
}
\begin{align}
\left(\;{\tiny\yng(1)}\;;\; 0\;\right)=\psi_0\;,\qquad  \left(\;{\tiny\yng(1)}\;;\; {\tiny\yng(1)}\;\right)=\omega_1\;,\qquad \left(\;0\;;\;{\tiny\yng(1)}\;\right)=\widetilde{\psi}_1
\end{align}
with conformal dimension $h_+=\frac{1}{2}\left(1+\lambda\right)\;,\;h_1\approx\frac{\lambda^2}{2N}$ and $h_-=\frac{1}{2}\left(1-\lambda\right)$, respectively.
Chang and Yin consider the  three point functions.
\begin{subequations}
\begin{align}
\begin{split}
C_{3}&\left((\;\overline{{\tiny\yng(1)}}\;;\;\overline{{\tiny\yng(1)}}\;),(\;\overline{{\tiny\yng(1)}}\;;\;\overline{{\tiny\yng(1)}}\;),(\;{\tiny\yng(1,1)}\;;\;{\tiny\yng(1,1)}\;)\right)\\
&=1-\frac{\lambda^2}{2N^2}\left(\pi\cot\pi\lambda-\pi^2\lambda\csc^2\pi \lambda+2\gamma+2\psi(\lambda)+2\lambda\psi^{(1)}(\lambda)\right)+\mathcal{O}\left(\frac{1}{N^3}\right)
\end{split}\\
\begin{split}
C_{3}&\left((\;\overline{{\tiny\yng(1)}}\;;\;\overline{{\tiny\yng(1)}}\;),(\;\overline{{\tiny\yng(1)}}\;;\;\overline{{\tiny\yng(1)}}\;),(\;{\tiny\yng(2)}\;;\;{\tiny\yng(2)}\;)\right)\\
&=1+\frac{\lambda^2}{2N^2}\left(\pi\cot\pi\lambda-\pi^2\lambda\csc^2\pi \lambda+2\gamma+2\psi(\lambda)+2\lambda\psi^{(1)}(\lambda)\right)+\mathcal{O}\left(\frac{1}{N^3}\right)
\end{split}
\end{align}
\end{subequations}
where $\left(\;\overline{{\tiny\yng(1)}}\;;\;\overline{{\tiny\yng(1)}}\;\right)$ denotes conjugate representation of $\left(\;{\tiny\yng(1)}\;;\; {\tiny\yng(1)}\;\right)=\omega_1$.
Based on this one obtains the relation:
\begin{align}
\frac{1}{\sqrt{2}}\left\{\left(\;{\tiny\yng(2)}\;;\;{\tiny\yng(2)}\;\right)+\left(\;{\tiny\yng(1,1)}\;;\;{\tiny\yng(1,1)}\;\right)\right\}=\frac{1}{\sqrt{2}}\omega_1^2
\end{align}
The orthogonal linear combination has a vanishing three point function in the large $N$ limit so one identifies
\begin{align}
\frac{1}{\sqrt{2}}\left\{\left(\;{\tiny\yng(2)}\;;\;{\tiny\yng(2)}\;\right)-\left(\;{\tiny\yng(1,1)}\;;\;{\tiny\yng(1,1)}\;\right)\right\}=\omega_2
\end{align}
as a new single trace operator.
In this way, Chang and Yin produced the following examples \cite{yin2011,yin2012}.
\begin{subequations}
\begin{align}
\omega_1\sim&\left(\;{\tiny\yng(1)}\;;\;{\tiny\yng(1)}\;\right)\label{identification_ex5}\\
\omega_2\sim&\left(\;{\tiny\yng(2)}\;;\;{\tiny\yng(2)}\;\right)-\left(\;{\tiny\yng(1,1)}\;;\;{\tiny\yng(1,1)}\;\right)\label{identification_ex6}\\
\omega_3\sim&\left(\;{\tiny\yng(3)}\;;\;{\tiny\yng(3)}\;\right)-\left(\;{\tiny\yng(2,1)}\;;\;{\tiny\yng(2,1)}\;\right)+\left(\;{\tiny\yng(1,1,1)}\;;\;{\tiny\yng(1,1,1)}\;\right)\label{identification_ex7}\\
\psi_2\sim&\sqrt{2}\left(\;{\tiny\yng(3)}\;;\;{\tiny\yng(2)}\;\right)-\left(\;{\tiny\yng(2,1)}\;;\;{\tiny\yng(2)}\;\right)-\left(\;{\tiny\yng(2,1)}\;;\;{\tiny\yng(1,1)}\;\right)+\sqrt{2}\left(\;{\tiny\yng(1,1,1)}\;;\;{\tiny\yng(1,1)}\;\right)
\end{align}
\end{subequations}
\begin{subequations}\label{identification_ex}
\begin{align}
\left(\;{\tiny\yng(1,1)}\;;\;0\right)=&\frac{1}{\sqrt{2}}\psi_0^2\\
\left(\;\left({\tiny\yng(1)},\overline{{\tiny\yng(1)}}\right)\;;\;0\right)=&\psi_0\overline{\psi}_0\\
\left(\;{\tiny\yng(1,1)}\;;\;{\tiny\yng(1,1)}\;\right)=&\frac{1}{2}\omega_1^2-\frac{1}{\sqrt{2}}\omega_2\label{identification_ex1}\\
\left(\;{\tiny\yng(2)}\;;\;{\tiny\yng(2)}\;\right)=&\frac{1}{2}\omega_1^2+\frac{1}{\sqrt{2}}\omega_2\label{identification_ex2}\\
\left(\;{\tiny\yng(1,1)}\;;\;{\tiny\yng(1)}\;\right)=&\frac{1}{\sqrt{2}}\left(\psi_1+\psi_0\omega_1\right)\label{identification_ex3}\\
\left(\;{\tiny\yng(2)}\;;\;{\tiny\yng(1)}\;\right)=&\frac{1}{\sqrt{2}}\left(-\psi_1+\psi_0\omega_1\right)\label{identification_ex4}
\end{align}
\end{subequations}
for the first few single traces. However, in general it is not easy to pursue the general identifications between $\left(\Lambda_+;\Lambda_-\right)$ states and single-trace operators following this technique. We shall in what follows present another method capable of giving a complete identification of higher states as Fock space eigenstates and a Hamiltonian with degenerate perturbation theory.

\section{The Method}

In what follows we will present a proposal for the structure of normalized single and multi-trace operators  and proceed to establish its validity. Let us denote  two infinite sequences of single-trace operators visible from the study of Chang and Yin \cite{yin2011,yin2012,Chang:2013izp} as: $\psi_n,\overline{\psi}_n,\omega_{n},\overline{\omega}_n$. The subscripts of the fields will be referred to as the  winding number. In addition, $\overline{\psi}_n, \overline{\omega}_m$ is a conjugate field of $\psi_n,\omega_m$, respectively.
\begin{subequations}
\begin{alignat}{3}
\psi_0,\psi_1,\psi_2,\cdots,\qquad&&\omega_1,\omega_2,\cdots\\
\overline{\psi}_0,\overline{\psi}_1,\overline{\psi}_2,\cdots,\qquad&&\overline{\omega}_1,\overline{\omega}_2,\cdots
\end{alignat}
\end{subequations}

In our construction, we follow an analogy with the matrix-vector model.  It will be seen that this will be more than an analogy, the correspondence will turn out to hold nontrivially at the full dynamical level.
For comparison with the structures in the matrix -vector model, it will be  convenient to use $\gamma_n$ which are defined by multiplying $\omega_n$ by $\sqrt{n}$.
\begin{align}
\begin{split}
\gamma_n\equiv\sqrt{n}\omega_n,\qquad\overline{\gamma}_n\equiv\sqrt{n}\overline{\omega}_n\\
\gamma_1,\gamma_2,\cdots,\qquad\overline{\gamma}_1,\overline{\gamma}_2,\cdots
\end{split}
\end{align}
For example , with this normalization  $\gamma_n$ would be equivalent to  the matrix invariant variable $tr\left(U^n\right)$ in \cite{antal1992,antal1996,antal2013}. Similarly, the second set of operators is related to the vector singlets of the matrix-vector model.
To address the space of all primary operators (states) our main proposal is to work in the Fock space based on the set of single trace operators as field (creation-anihilation) variables. To make the problem tractable, we will first restrict the space of primary operators considered. In the present paper we will consider primaries without  derivatives; therefore, the field theory  constructed will correspond to certain localization of the full theory. In particular  we will work in the sector of the one scalar field ( suppressing states related to the second scalar $\tilde{\psi}$ ). We will show subsequently that there is a certain symmetry between the two scalar sequences so that an equivalent construction holds in the `mirror'  $\tilde{\psi}$ sector case. In Young tableaux $\left(\Lambda_+;\Lambda_-\right)$ notation, this will mean that we concentrate our attention on operators where the Young tableau $S_+,R_+$ has at most one more box than Young tableau $S_-,R_-$ in each row, respectively. When $S_+$ has two more boxes than $\Lambda_-$ in some row, then this state is related to derivatives of single-trace operators. For example, in \cite{yin2011}, we have
\begin{align}
\left(\;{\tiny\yng(2)}\;;\;0\;\right)=\frac{1}{\sqrt{2}\Delta_{(f,0)}}\left(\psi_0\partial\overline{\partial}\psi_0-\partial\psi_0\overline{\partial}\psi_0\right)
\end{align}
Thus, we will ignore these cases. Though the number of derivatives can be calculated simply, its specific structure is nontrivial, we plan to analyze this problem in  future study.

We also state a further simplifying caveat. The sequence of primaries labeled as `light states'  was shown to obey approximate conservation equations, involving both $\psi$'s and $\widetilde{\psi}$'s,  \cite{raju2011,yin2011} such as
\begin{align}
\frac{1}{\Delta_{\left({\tiny\yng(1)};{\tiny\yng(1)}\right)}}{\partial}\overline{\partial}\omega\sim \psi_0\widetilde{\psi}_0
\end{align}
 Once the extension of our construction to two scalar sequences $\tilde{\psi}$'s is given these type of equations are to be imposed as `Gauss law' conditions, a problem left for future work.

\subsection{Subsector $\left(\Lambda;\Lambda\right)$}
Consider first a subsector of states given by $\Lambda=\Lambda_+=\Lambda_-$, with the conformal dimensions given by
\begin{align}
h\left(\Lambda;\Lambda\right)=\frac{\frac{\lambda^2}{N^2}}{1+\frac{\lambda}{N}}\left(\frac{1}{2}BN+\frac{1}{2}D-\frac{1}{2N}B^2\right)=\frac{\frac{\lambda^2}{N^2}}{1+\frac{\lambda}{N}}C\left(\Lambda\right)
\end{align}
Our first observation is to recognize that Chang and Yin's results \eqref{identification_ex5}, \eqref{identification_ex6} and \eqref{identification_ex7} :
\begin{subequations}
\begin{align}
\omega_1\sim& P_1\left({\tiny\yng(1)};\{\sqrt{i}\omega_i\}\right)\\
\omega_2\sim& P_2\left({\tiny\yng(2)};\{\sqrt{i}\omega_i\}\right)-P_2\left({\tiny\yng(1,1)};\{\sqrt{i}\omega_i\}\right)\\
\omega_3\sim& P_3\left({\tiny\yng(3)};\{\sqrt{i}\omega_i\}\right)-P_3\left({\tiny\yng(2,1)};\{\sqrt{i}\omega_i\}\right)+P_3\left({\tiny\yng(1,1,1)};\{\sqrt{i}\omega_i\}\right)
\end{align}
\end{subequations}
take the form of  Schur polynomials. For example,
\begin{subequations}
\begin{align}
P_1\left({\tiny\yng(1)};\{x_i\}\right)&=x_1\\
P_2\left({\tiny\yng(2)};\{x_i\}\right)&=\frac{1}{2}x_1^2+\frac{1}{2}x_2\\
P_2\left({\tiny\yng(1,1)};\{x_i\}\right)&=\frac{1}{2}x_1^2-\frac{1}{2}x_2\\
P_3\left({\tiny\yng(3)};\{x_i\}\right)&=\frac{1}{6}x_1^3+\frac{1}{2}x_1x_2+\frac{1}{3}x_3\\
P_3\left({\tiny\yng(2,1)};\{x_i\}\right)&=\frac{1}{3}x_1^3-\frac{1}{3}x_3\\
P_3\left({\tiny\yng(1,1,1)};\{x_i\}\right)&=\frac{1}{6}x_1^3-\frac{1}{2}x_1x_2+\frac{1}{3}x_3
\end{align}
\end{subequations}

These appeared earlier \cite{antal1992}. In the matrix model collective Hamiltonian corresponding to a hamiltonian matrix model
\begin{align}
H=\underbrace{N\sum_{n=1}^{\infty}na_{n}^\dag a_n}_{\text{Quadratic terms}}+\underbrace{\frac{1}{2}\sum_{\substack{n,m>0\\n,m<0}}\sqrt{nm\left|n+m\right|}\left\{a_m^\dag a_{n}^\dag a_{n+m}+a_{n+m}^\dag a_m a_n\right\}}_{\text{Cubic terms}}
\end{align}
It was shown in \cite{antal1992} that the Fock space eigenstates  of H are Schur polynomials $P_n\left(\Lambda;\{\sqrt{i}a^\dag_i\}\right)$ with eigenvalues
\begin{align}
E_n\left(\Lambda\right)=2C\left(\Lambda\right)+\frac{\left|B\right|^2}{N}
\end{align}

There is also a more general construction based on characters of $U(N)$ or CFT which we describe in a separate publication \cite{antal2013}.
Therefore, to describe  interactions between $\omega$' states we postulate the identical collective Hamiltonian 
\begin{align}
\label{cubicinteraction}
\begin{split}
\frac{\frac{\lambda^2}{N^2}}{1+\frac{\lambda}{N}}&\left[\frac{N}{2}\sum_{n=1}^\infty n\omega_n\frac{\partial}{\partial \omega_n}-\frac{1}{2N}\left(\sum_{n=1}^\infty n\omega_n\frac{\partial}{\partial \omega_n}\right)^2+\frac{1}{2}\sum_{n=2}^\infty\sum_{m=1}^{n-1}\sqrt{nm(n-m)}\omega_m\omega_{n-m}\frac{\partial}{\partial \omega_n}\right. \\
&\left.+\frac{1}{2}\sum_{n=1}^\infty\sum_{m=1}^{\infty}\sqrt{nm(n+m)}\omega_{n+m}\frac{\partial^2}{\partial \omega_n\partial \omega_m}\right]
\end{split}
\end{align}

According to \cite{antal1992}, eigenstates of this Hamiltonian are Schur polynomials of $\gamma_n=\sqrt{n}\omega_n$\footnote{For explicit form, see \eqref{lightstate}}. From the completeness relation of characters of $S_n$, we can express single-trace operator $\omega_n$ or $\gamma_n$ in terms of $\left(S;S\right)$ where $S\in \text{rep}\left(S_n\right)$.
\begin{align}
\begin{split}
\sum_{S\in\text{rep}\left(S_n\right)}ch_S(g)\left(S;S\right)=\prod_{i=1}^\infty\left(\gamma_i\right)^{\lambda(g)_i}\qquad\mbox{where}\quad g\in S_n
\end{split}
\end{align}
where rep$\left(S_n\right)$ is a set of all irreducible representations of $S_n$, which correspond to all Young tableaux of $n$ boxes.  

Especially, for $g=g^{\left(0\right)}\equiv\left(1,2,\cdots,n\right)$ which corresponds to a column of $n$ boxes\footnote{e.g. ${\tiny\yng(1,1)},\;{\tiny\yng(1,1,1)},\;{\tiny\yng(1,1,1,1)},\cdots$},
\begin{align}
\omega_n=\frac{1}{\sqrt{n}}\gamma_n=\frac{1}{\sqrt{n}}\sum_{S\in\text{rep}\left(S_n\right)}ch_S\left(g^{\left(0\right)}\right)\left(S;S\right)
\end{align}
By using basic knowledge of the permutation group, we can calculate the coefficients explicitly.
\begin{align}
ch_{\Lambda}\left(g^{\left(0\right)}\right)=\begin{cases}\left(-1\right)^{k+1}&\text{where}\quad \Lambda=\left(n-k\right)w_1+w_{k}\quad\left(k=1,2,\cdots,n\right)\\
0&\text{otherwise}\\
\end{cases}
\end{align}
The condition of $\Lambda$ for non-zero character corresponds to 
\begin{align}
r_i=\begin{cases}
n-k+1&,i=1\\
1&,i=2,\cdots, k\\
0&,i=k+1,\cdots,n\\
\end{cases}
\qquad \left(k=1,2,\cdots,n\right)
\end{align}
For example, for $\omega_3$,
\begin{align}
ch_{{\tiny\yng(3)}}\left(g^{\left(0\right)}\right)=1,\qquad ch_{{\tiny\yng(2,1)}}\left(g^{\left(0\right)}\right)=-1,\qquad ch_{{\tiny\yng(1,1,1)}}\left(g^{\left(0\right)}\right)=1,\qquad\mbox{others}=0
\end{align}
and, we get
\begin{align}
\begin{split}
\omega_3=\frac{1}{\sqrt{3}}\left[\left(\;{\tiny\yng(3)}\;;\;{\tiny\yng(3)}\;\right)-\left(\;{\tiny\yng(2,1)}\;;\;{\tiny\yng(2,1)}\;\right)+\left(\;{\tiny\yng(1,1,1)}\;;\;{\tiny\yng(1,1,1)}\;\right)\right]
\end{split}
\end{align}
This agrees with the result of \cite{yin2012}. We can generate other results immediately. For instance,
\begin{align}
\begin{split}
\omega_4=\frac{1}{\sqrt{4}}\left[\left(\;{\tiny\yng(4)}\;;\;{\tiny\yng(4)}\;\right)-\left(\;{\tiny\yng(3,1)}\;;\;{\tiny\yng(3,1)}\;\right)+\left(\;{\tiny\yng(2,1,1)}\;;\;{\tiny\yng(2,1,1)}\;\right)-\left(\;{\tiny\yng(1,1,1,1)}\;;\;{\tiny\yng(1,1,1,1)}\;\right)\right]
\end{split}
\end{align}

As in \cite{antal1992} this Hamiltonian with the cubic interaction preserves the winding numbers, which is used to classify its eigenstates. This suggests that the extra (winding number) dimension  will  play an important role in our full construction.

\subsection{Extension}
A central role which emerges in the construction is the appearance of an extra `winding' mode coordinate . Such extra dimensions appear naturally in the matrix-vector model framework \cite{antal1996}. In \cite{antal1996}, $\psi_n$ as a winding number:
\begin{align}
\psi_i\sim \overline{x}U^i x\;,\qquad\omega_j\sim tr\left( U^j\right)\label{matrixvectorvariables}
\end{align}
where $x$ is a vector and $U$ is a matrix. With this analogy  we define the winding number of $\psi_i$, $\omega_j$ to be $i$, $j$, respectively. From data of three point functions, we will  establish a collective Hamiltonian which will be seen to preserve the total winding number of $\psi$'a and $\omega$'s. The two  winding number operators are given as:
\begin{subequations}
\begin{align}
K=\sum_{i=0}^\infty i\psi_i\frac{\partial}{\partial \psi_i}+\sum_{i=1}^\infty i \omega_i\frac{\partial}{\partial \omega_i}\;,\qquad\overline{K}=\sum_{i=0}^\infty i\overline{\psi}_i\frac{\partial}{\partial \overline{\psi}_i}+\sum_{i=1}^\infty i \overline{\omega}_i\frac{\partial}{\partial \overline{\omega}_i}
\end{align}
\end{subequations}
They will represent conserved quantities with the collective field Hamiltonian commuting with  with $K$ and $\overline{K}$.

Looking at the exact CFT expression for the  conformal dimension of primaries , we can get more information regarding the structure of the full Hamiltonian. The conformal dimension\footnote{For simplicity, we will consider $R_\pm=0$ cases. The extension to general case is straightforward and gives the same result.} is 
\begin{align}
h\left(\Lambda_+;\Lambda_-\right)=\frac{\lambda}{2}\left(B_+-B_-\right)+\frac{1}{2}\sum_{i=1}^{N-1} \left(r_{i}^+-r_{i}^-\right)^2+\mathcal{O}\left(\frac{1}{N}\right)
\end{align}
Since we are interested in states where $\Lambda_+$ has at most one more box than $\Lambda_-$ in each row, 
\begin{align}
h\left(\Lambda_+;\Lambda_-\right)=\frac{1}{2}\left(1+\lambda\right)\left(B_+-B_-\right)+\mathcal{O}\left(\frac{1}{N}\right)
\end{align}
Define a subspace $Y_{s_+,s_-;r_+,r_-}$ such that
\begin{align}
Y_{s_+,s_-;r_+,r_-}\equiv\left\{\left(\Lambda_+;\Lambda_-\right)\right|\left.\mbox{under the following three conditions}\right\}
\end{align}
\begin{enumerate}[itemsep=0mm]
\item $\left(\Lambda_+;\Lambda_-\right)=(\overline{R}_+,S_+;\overline{R}_-,S_-)$\\
    where  $r_+=\left|R_+\right|,s_+=\left|S_+\right|,r_-=\left|R_-\right|,s_-=\left|S_-\right|$
\item $S_-,R_-$ are sub-Young tableaux of $S_+,R_+$, respectively.
\item In every row, $R_+,S_+$ has at most one more box than $R_-,S_-$, respectively.
\end{enumerate}
Then, all states in $Y_{s_+,s_-;r_+,r_-}$ with fixed $r_\pm,s_\pm$ have the same conformal dimension up to order $\mathcal{O}\left(1\right)$. Moreover, the conformal dimension of $\left(\Lambda,\Lambda\right)$, which is composed of only $\omega$'s, is
\begin{align}
h\left(\Lambda;\Lambda\right)=\mathcal{O}\left(\frac{1}{N}\right)
\end{align}
Thus, the contribution of order $\mathcal{O}\left(1\right)$ to the conformal dimension, which depends on $s_+-s_-$ and $r_+-r_-$, does not come from $\omega$'s, but from $\psi$'s. In \cite{antal1992}, the quadratic term is unperturbed Hamiltonian and the cubic interaction term corresponds to perturbation. Hence, $\prod_{n}a_n^\dag\left|0\right>$ is an eigenstate of unperturbed Hamiltonian, but it is not eigenstate of full Hamiltonian. Likewise, we expect that $\psi$'s and $\omega$'s are eigenstates of unperturbed Hamiltonian corresponding to eigenvalues, $\frac{1}{2}\left(1+\lambda\right)\left(B_+-B_-\right)$ up to order $\mathcal{O}\left(1\right)$. \eqref{identification_ex3} and \eqref{identification_ex4} imply that the eigenvalues of unperturbed Hamiltonian of both $\psi_0$ and $\psi_1$ are $\frac{1}{2}\left(1+\lambda\right)$.
 
From this observation, we can assume
\begin{align}
H_{\text{unperturbed}}\psi_i=\left[\frac{1}{2}\left(1+\lambda\right)+\mathcal{O}\left(\frac{1}{N}\right)\right]\psi_i\qquad \mbox{for all }i=0,1,2,\cdots
\end{align}
Then,
\begin{align}
Y_{s_+,s_-;0,0}=\left\{\sum\left(\prod_{i=1}^{s_+-s_-}\psi_{a_i}\prod_{j}\omega_{b_j}\right)\right\}
\end{align}
where we consider only the case where $R_\pm=0$ for simplicity. The extension to all cases are straightforward. Note that the number of $\psi$'s of states in $Y_{s_+,s_-;0,0}$ is equal to $\left|S_+\right|-\left|S_-\right|$. 

Define operators $M,\overline{M}$ which commute with the collective field Hamiltonian.
\begin{align}
M=\sum_{i=0}^\infty \psi_i\frac{\partial}{\partial \psi_i},\qquad\overline{M}=\sum_{i=0}^\infty \overline{\psi}_i\frac{\partial}{\partial \overline{\psi}_i}
\end{align}
This corresponds to the total number of $\psi$'s and $\overline{\psi}$'s, respectively.

Moreover, considering \eqref{identification_ex}, each term of eigenstate $(S_+;S_-)$ has the same total winding number, which seems to be equal to $\left|S_-\right|$. Thus, we add one more assumption that the total winding number is equal to $\left|S_-\right|$.

\subsection{Counting Argument}

The assumptions in the previous section come from observing several examples of small Young tableaux. The full agreement can be established as follows

Consider the subspace subspace $F^{k,\overline{k}}_{m,\overline{m}}$ such that
\begin{align}
F_{m,\overline{m}}^{k,\overline{k}}=\left\{\prod_{i=1}^{m}\psi_{a_i}\prod_{j}\omega_{b_j}\prod_{i=1}^{\overline{m}}\overline{\psi}_{\overline{a}_i}\prod_{j}\overline{\omega}_{\overline{b}_j}\right|\left.\sum_{i=1}^{m}a_i+\sum_j b_j=k,\quad \sum_{i=1}^{\overline{m}}\overline{a}_i+\sum_j \overline{b}_j=\overline{k}\right\}
\end{align}
where $a_i,\overline{a}_i, b_i,\overline{b}_i$ are non-negative integers. Then, we will show that there is a one-to-one correspondence between the Young tableau states and these Fock spaces.
\begin{align}
\left|Y_{s_+,s_-;r_+,r_-}\right|=\left|F^{k,\overline{k}}_{m,\overline{m}}\right|
\end{align}
where
\begin{align*}
m&=s_+-s_-,\qquad k=s_-\\
\overline{m}&=r_+-r_-,\qquad \overline{k}=r_-
\end{align*}
For example,
\begin{align}
\left|\left\{\left(\;{\tiny\yng(2)}\;;\;{\tiny\yng(1)}\;\right),\left(\;{\tiny\yng(1,1)}\;;\;{\tiny\yng(1)}\;\right)\right\}\right|=\left|\left\{\psi_1,\psi_0\omega_1\right\}\right|=2
\end{align}

This means that the number of states in the eigenspace of unperturbed Hamiltonian (e.g. $\psi_1,\psi_0\omega_1$) is equal to that in the corresponding split eigenstates of the full Hamiltonian (e.g. $\left(\;{\tiny\yng(2)}\;;\;{\tiny\yng(1)}\;\right),\left(\;{\tiny\yng(1,1)}\;;\;{\tiny\yng(1)}\;\right)$). This is a non-trivial agreement that supports our construction .

First of all, we will prove the simplest case $Y_{k+1,k;0,0}$ where $S_+$ has one more box than $S_-$ with $\left|S_-\right|=k$ and $R_\pm=0$, which can be translated into $F^{k,0}_{1,0}$.

The number of states in $\left|F^{k,0}_{1,0}\right|$ can be easily calculated. A state in $F^{k,0}_{1,0}$ is
\begin{align}
\psi_a\prod_j \omega_{b_j}\in F_{1,0}^{k,0}\qquad\mbox{where}\quad\sum_j b_j=k-a
\end{align}
Since $a=0,1,\cdots, k$,
\begin{align}
\left|F_{1,0}^{k,0}\right|=\sum_{i=0}^k p\left(i\right)
\end{align}
where $p(x)$ is the number of partitions of $x$.

On the other hand, for the number of states in $Y_{k+1,k;0,0}$, we need to find a map from $Y_{k+1,k;0,0}$ to a partition of the number $k$. A state in $Y_{k+1,k;0,0}$ is
\begin{align}
\left(\Lambda_+;\Lambda_-\right)=\left(S_+;S_-\right)\quad, \qquad\mbox{where } \left|S_+\right|=k+1,\;\left|S_-\right|=k
\end{align}

The number of these states is equal to the number of ways to add one box to all possible $S_-$'s with $k$ boxes. Recalling that a Young tableau can be represented by partition of an integer, this counting problem can be modified into a problem of counting partitions. For example, we consider how to count the way to add one box to Young tableau ${\tiny\yng(2,1)}$ and see how this corresponds to the transformation of the corresponding partition of 3.
\begin{align}
\begin{array}{ccc}
{\young(aa,a)}&\longleftrightarrow&3=2+1\\
\downarrow&&\downarrow\\
{\young(aab,a)}&\longleftrightarrow&4=2+1+1
\end{array}
\end{align}
First of all, we can add one box (box b) at the first row. This corresponds to adding ``$+1$" to the original partition of 3.
\begin{align}
\begin{array}{ccc}
{\young(aa,a)}&\longleftrightarrow&3=2+1\\
\downarrow&&\downarrow\\
{\young(aa,ab)}&\longleftrightarrow&4=2+2
\end{array}
\end{align}
The next possible way is to add the box b to the second row. This corresponds to changing ``$+1$" in the original partition of 3 into ``$+2$".

In general, addition of one box at the $i$th rows corresponds to changing ``$+(i-1)$" of the partition into ``$+i$". Thus, the number of states is equal to the number of possible ways of these actions on all possible partitions of $k$.

Alternatively, we can count the number of these actions in a different way. First, we can add ``$+1$" to all partitions of $k$. Then, this number is equal to the number of all partitions of $k$, $p(k)$. Next, among all partitions, we can find partitions which have ``$+1$" and we can change this ``$+1$" into ``$+2$". The number of this action is equal to the number of partitions which contain ``$+1$". Thus, it is $p(k-1)$. In this way, we can conclude that the number of all states in $Y_{k+1,k;0,0}$ is
\begin{align}
\left|Y_{k+1,k;0,0}\right|=\sum_{i=0}^k p\left(i\right)=\left|F^{k,0}_{1,0}\right|
\end{align}

So far, we have proven our claim for the case where $S_+$ has one more box than $S_-$. In general, $S_+$ can have one or more than two boxes than $S_-$. In the appendix~\ref{appendixcountingstates}, we prove this general case. In fact, we only consider $R_\pm=0$ case. However, two sets of Young tableaux, $S_\pm$ and $R_\pm$ are almost independent so that the extension is straightforward. Thus, we can get
\begin{align}
\left|Y_{s_+,s_-;r_+,r_-}\right|=\left|F^{k,\overline{k}}_{m,\overline{m}}\right|
\end{align}

Concluding this analysis we state the properties of the Fock space theory :
\begin{enumerate}[itemsep=-1mm]
\item Conservation of the total number of $\psi$'s

\item Conservation of the total winding number

\item $\left(\Lambda_+;\Lambda_-\right)$ is eigenstate of the full Hamiltonian while $\psi$'s and $\omega$'s are eigenstates of unperturbed Hamiltonian.

\item These properties are same for $\overline{\psi}$'s and $\overline{\omega}$'s

\item The perturbative coupling contants are $\frac{\lambda}{N}$ and $\epsilon=\frac{\frac{\lambda^2}{N^2}}{1+\frac{\lambda}{N}}$. (see \eqref{conformaldimension})
\end{enumerate}

\section{The Hamiltonian}

From the structure of Fock space established in the previous analysis we are led to the following form of the collective field Hamiltonian:
\begin{align}
\begin{split}
L_0=&\overbracket{\frac{1}{2}\left(1+\lambda\right)\left(M+\overline{M}\right)+Q\left(M,\overline{M},K,\overline{K}\right)}^{\text{unperturbed Hamiltonian}}\\
&+\epsilon\left[ \frac{1}{2}\sum_{n=2}^\infty\sum_{m=1}^{n-1}\sqrt{nm(n-m)}\omega_m\omega_{n-m}\frac{\partial}{\partial \omega_n} +\frac{1}{2}\sum_{n=1}^\infty\sum_{m=1}^{\infty}\sqrt{nm(n+m)}\omega_{n+m}\frac{\partial^2}{\partial \omega_n\partial \omega_m}\right]\\
&+\frac{\lambda}{N}\left[\sum_{n,m}A(n,m)\psi_{n+m}\frac{\partial^2}{\partial \psi_n\partial \omega_m}+\sum_{n,m} B(n,m)\psi_{n-m}\omega_m\frac{\partial}{\partial \psi_n}\right]\\
&+\epsilon\left[\sum_{n,m}C(n,m)\psi_{n+m}\frac{\partial^2}{\partial \psi_n\partial \omega_m}+\sum_{n,m} D(n,m)\psi_{n-m}\omega_m\frac{\partial}{\partial \psi_n}\right]\\
&+\frac{\lambda}{N}\sum_{n,m,u}F(n,m,u)\psi_{n+m-u}\psi_u\frac{\partial^2}{\partial \psi_n\partial \psi_m}+\epsilon\sum_{n,m,u}G(n,m,u)\psi_{n+m-u}\psi_u\frac{\partial^2}{\partial \psi_n\partial \psi_m}
\end{split}
\end{align}
Here  
 $Q\left(M,\overline{M},K,\overline{K}\right)$ is a function of global operators $M,\overline{M},K,\overline{K}$ and  $A(n,m)$, $B(n,m)$, $C(n,m)$, $D(n,m)$, $F(n,m,u)$ and $G(n,m,u)$ are still the most general form factors whose precise form we will establish shortly

\subsection{Determining Coupling Constant from Three Point Functions}

Having the general form of the Hamiltonian, the next task is to determine its coefficients (form factors). These can be determined by precise comparison  between $Y_{s_+,s_-;r_+,r_-}$ and $F^{k,\overline{k}}_{m,\overline{m}}$ spaces , which can be obtained from the knowledge of  three point functions. We have  already seen  the linear transformation between $Y_{2,1;0,0}$ and $F^{1,0}_{1,0}$ in \eqref{identification_ex3} and \eqref{identification_ex4}. The next example is $Y_{3,2;0,0}$ and $F^{2,0}_{1,0}$. Conisder
\begin{align}
\begin{pmatrix}
\left(\;{\tiny\yng(1,1,1)}\;;\;{\tiny\yng(1,1)}\;\right)\\
\left(\;{\tiny\yng(2,1)}\;;\;{\tiny\yng(2)}\;\right)\\
\left(\;{\tiny\yng(2,1)}\;;\;{\tiny\yng(1,1)}\;\right)\\
\left(\;{\tiny\yng(3)}\;;\;{\tiny\yng(2)}\;\right)
\end{pmatrix}=A\begin{pmatrix}
\psi_2\\
\psi_1\omega_1\\
\psi_0\omega_1^2\\
\psi_0\omega_2
\end{pmatrix}=\begin{pmatrix}
A_{1,1} & A_{1,2} & A_{1,3} & A_{1,4}\\
A_{2,1} & A_{2,2} & A_{2,3} & A_{2,4}\\
A_{3,1} & A_{3,2} & A_{3,3} & A_{3,4}\\
A_{4,1} & A_{4,2} & A_{4,3} & A_{4,4}\\
\end{pmatrix}\begin{pmatrix}
\psi_2\\
\psi_1\omega_1\\
\psi_0\omega_1^2\\
\psi_0\omega_2
\end{pmatrix}
\end{align}
where $A$ is a constant $4\times 4$ matrix. Due to practical difficulty\footnote{In \cite{yin2011}, the formula for three point function has infinite products in the large $N$ limit.}, we can calculate a few of the three point functions to determine elements of $A$. For example, we can calculate
\begin{align}
C_3&\left((\;\overline{{\tiny\yng(1)}}\; ;\; 0\;),(\;\overline{{\tiny\yng(2)}}\; ;\; \overline{{\tiny\yng(2)}} \;),(\;{\tiny\yng(2,1)}\; ;\;{\tiny\yng(2)}\;)\right)=\sqrt{\frac{2}{3}}+\mathcal{O}\left(\frac{1}{N}\right)\label{threepointfunctionresult1}
\end{align}
but, some three point functions would be hard.\footnote{This calculation is not impossible. In fact, we derived finite products from the formula of \cite{yin2011}. And, this derivation is valid for special cases. For example, we can apply this reduced formula to \eqref{threepointfunctionresult1}, but cannot to \eqref{threepointfunctionnot}} For instance, 
\begin{align}
C_3&\left((\;\overline{{\tiny\yng(1)}}\; ;\; 0\;),(\;\overline{{\tiny\yng(1,1)}}\; ;\; \overline{{\tiny\yng(1,1)}} \;),(\;{\tiny\yng(2,1)}\; ;\;{\tiny\yng(2)}\;)\right)\label{threepointfunctionnot}
\end{align}

Nevertheless, we may get the answer by assuming symmetry in three point functions. In appendix~\ref{threepointfunctionappendix} and \cite{yin2011}, the leading order in structure constants of the three point functions seem to be invariant under transpose.
\begin{align}
C_3\left(\left(R_1;R_2\right),\left(R_3;R_4\right),\left(R_5;R_6\right)\right)\simeq C_3\left(\left(R_1^t;R_2^t\right),\left(R_3^t;R_4^t\right),\left(R_5^t;R_6^t\right)\right)
\end{align}
up to order $\mathcal{O}\left(1\right)$.
By using this symmetry, we can get
\begin{align}
C_3\left((\;\overline{{\tiny\yng(1)}}\; ;\; 0\;),(\;\overline{{\tiny\yng(1,1)}}\; ;\; \overline{{\tiny\yng(1,1)}} \;),(\;{\tiny\yng(2,1)}\; ;\;{\tiny\yng(2)}\;)\right)=C_3\left((\;\overline{{\tiny\yng(1)}}\; ;\; 0\;),(\;\overline{{\tiny\yng(2)}}\; ;\; \overline{{\tiny\yng(2)}} \;),(\;{\tiny\yng(2,1)}\; ;\;{\tiny\yng(1,1)}\;)\right)=0\label{threepointfunctionresult2}
\end{align}
Combining \eqref{threepointfunctionresult1} and \eqref{threepointfunctionresult2}, we get
\begin{align}
A_{2,3}=\frac{1}{\sqrt{6}},\qquad A_{2,4}=\frac{1}{\sqrt{3}}
\end{align}
In the same way for $\left(\;{\tiny\yng(1,1,1)}\;;\;{\tiny\yng(1,1)}\;\right)$, $\left(\;{\tiny\yng(2,1)}\;;\;{\tiny\yng(1,1)}\;\right)$ and $\left(\;{\tiny\yng(3)}\;;\;{\tiny\yng(2)}\;\right)$, 
\begin{align}
\begin{split}
A_{1,3}=A_{4,3}=\frac{1}{2\sqrt{3}},\qquad A_{3,3}=A_{3,4}=-A_{1,4}=\frac{1}{\sqrt{6}},\qquad A_{3,4}=-\frac{1}{\sqrt{3}}
\end{split}
\end{align}
Moreover, considering $C_3\left((\;\overline{{\tiny\yng(1,1)}}; ;\; \overline{{\tiny\yng(1)}}\;),(\;\overline{{\tiny\yng(1)}}\; ;\; \overline{{\tiny\yng(1)}} \;),(\;{\tiny\yng(2,1)}\; ;\;{\tiny\yng(2)}\;)\right)$, $C_3\left((\;\overline{{\tiny\yng(1,1)}}; ;\; \overline{{\tiny\yng(1)}}\;),(\;\overline{{\tiny\yng(1)}}\; ;\; \overline{{\tiny\yng(1)}} \;),(\;{\tiny\yng(2,1)}\; ;\;{\tiny\yng(1,1)}\;)\right)$ and their transpose three point functions, we can further determine
\begin{align}
A_{2,2}=-\frac{1}{\sqrt{6}},\qquad A_{3,2}=\frac{1}{\sqrt{6}}
\end{align}

Other coefficients can be fixed by normalization condition and transpose symmetry up to sign. A difference choice of sign (especially, sign of $A_{2,1}$ and $A_{3,1}$) will only change signs of coefficient of collective field Hamiltonian. We fixed the sign such that the sign of coefficients in collective field Hamiltonian is equal to the that of result in \cite{antal1996}. This will be shown in section \ref{connectiontomv}.

The final result is
\begin{align}
\begin{pmatrix}
\left(\;{\tiny\yng(1,1,1)}\;;\;{\tiny\yng(1,1)}\;\right)\\
\left(\;{\tiny\yng(2,1)}\;;\;{\tiny\yng(2)}\;\right)\\
\left(\;{\tiny\yng(2,1)}\;;\;{\tiny\yng(1,1)}\;\right)\\
\left(\;{\tiny\yng(3)}\;;\;{\tiny\yng(2)}\;\right)
\end{pmatrix}=\begin{pmatrix}
\frac{1}{\sqrt{3}}&-\frac{1}{\sqrt{3}}&\frac{1}{2\sqrt{3}}&-\frac{1}{\sqrt{6}}\\
-\frac{1}{\sqrt{6}}&-\frac{1}{\sqrt{6}}&\frac{1}{\sqrt{6}}&\frac{1}{\sqrt{3}}\\
-\frac{1}{\sqrt{6}}&\frac{1}{\sqrt{6}}&\frac{1}{\sqrt{6}}&-\frac{1}{\sqrt{3}}\\
\frac{1}{\sqrt{3}}&\frac{1}{\sqrt{3}}&\frac{1}{2\sqrt{3}}&\frac{1}{\sqrt{6}}\\
\end{pmatrix}\begin{pmatrix}
\psi_2\\
\psi_1\omega_1\\
\psi_0\omega_1^2\\
\psi_0\omega_2
\end{pmatrix}
\end{align}

Now, we act collective field Hamiltonian $L_0$ on $\{\psi_2,\psi_1\omega_1, \frac{1}{\sqrt{2}}\psi_0\omega_1^2,\psi_0,\omega_2\}$. Because $\left(\Lambda_+;\Lambda_-\right)$ is an eigenstate of collective field Hamiltonian and its eigenvalue is the corresponding conformal dimension, the Hamiltonian can be represented in the basis of $\{\psi_2,\psi_1\omega_1, \frac{1}{\sqrt{2}}\psi_0\omega_1^2,\psi_0,\omega_2\}$.
\begin{align}
L_0=\begin{pmatrix}
E_{3,2}&\epsilon+\frac{\lambda}{N}&0&\sqrt{2}\frac{\lambda}{N}\\
\epsilon+\frac{\lambda}{N}&E_{3,2}&\sqrt{2}\frac{\lambda}{N}&0\\
0&\sqrt{2}\frac{\lambda}{N}&E_{3,2}&\epsilon\\
\sqrt{2}\frac{\lambda}{N}&0&\epsilon&E_{3,2}\\
\end{pmatrix}
\end{align}
where $E_{3,2}=\frac{1}{2}\left(1+\lambda\right)-\frac{1}{2N}-\frac{5\lambda}{2N^2}+\epsilon\left(N-\frac{2}{N}\right)$ and $\epsilon=\frac{\frac{\lambda^2}{N^2}}{1+\frac{\lambda}{N}}$. From this result, we can determine a few coefficients $A,B,C,D$ in the collective Hamiltonian.
\begin{alignat*}{4}
A(0,1)=1&&,\quad A(0,2)=&&\sqrt{2},\quad A(1,1)=&&1\\
B(1,1)=1&&,\quad B(2,1)=&&1,\quad B(2,2)=&&\sqrt{2}\\
C(0,1)=0&&,\quad C(0,2)=&&0,\quad C(1,1)=&&1\\
D(1,1)=0&&,\quad D(2,1)=&&1,\quad D(2,2)=&&0
\end{alignat*}

\subsection{Determining  the Coupling Constants  by Diagonalizing the Hamiltonian}

For coefficients $A,B,C,D$ of larger $n,m$, we have to analyze $Y_{s+1,s;0,0}$ ($s\geqq 3$) in the same manner so that we can ignore interactions between $\psi$'s. However, calculation of three point functions is not easy in these cases. Instead, we may diagonalize collective Hamiltonian directly. However, the collective Hamiltonian could be diagonalized in the subspace $F^{k,\overline{k}}_{m,\overline{m}}$ with any coefficients $A,B,C,D$. Nevertheless, if corresponding eigenvalues are equal to the conformal dimension and if corresponding degeneracies -if any- are equal to $W_N$ minimal model, then such correspondence will not be a mere accident.

We will diagonalize the collective field Hamiltonian in the subspace $Y_{4,3;0,0}=F^{3,0}_{1,0}$. We can represent collective field Hamiltonian in the basis of $F^{3,0}_{1,0}$ with undetermined variables, $A, B, C$ and $D$. Then, when we diagonalize this matrix, we want the corresponding eigenvalues to be the conformal dimensions of states in $Y_{4,3;0,0}$. Especially, since we know all conformal dimensions of $Y_{4,3;0,0}$, we can calculate a characteristic polynomial either from the matrix directly or from the eigenvalues which are expected to be the conformal dimensions. By comparing coefficients of the characteristic polynomial from both of them, we can fix $A,B,C,D$. The result is
\begin{alignat*}{4}
A(0,3)=&&\sqrt{3},\quad A(1,2)=&&\sqrt{2},\quad A(2,1)=&&1\\
B(3,1)=&&1,\quad B(3,2)=&&\sqrt{2},\quad B(3,3)=&&\sqrt{3}\\
C(0,3)=&&0,\quad C(1,2)=&&\sqrt{2},\quad C(2,1)=&&2\\
D(3,1)=&&2,\quad D(3,2)=&&\sqrt{2},\quad D(3,3)=&&0
\end{alignat*}
From above all data, we may guess
\begin{align}
A(n,m)=B(n,m)=\sqrt{m},\qquad C(n,m)=n\sqrt{m},\qquad D(n,m)=(n-m)\sqrt{m}
\end{align}
Using this guess, we can calculate representation of the collective Hamiltonian in the basis of $F^{4,0}_{1,0}$. The eigenvalues of this matrix are exactly the same as the conformal dimensions of states in $Y_{5,4;0,0}$. The detailed result is in section~\ref{eigenstates}. 

We still need to determine coefficients $F,G$. In order to fix them, we have to consider $Y_{k+m,k;0,0}=F^{k,0}_{m,0}$ for $m\geqq 2$. In the same way, we obtained several $F(n,m,u)$ and $G(n,m,u)$ from the following subspaces.
\begin{align*}
Y_{3,1;0,0}&=F^{1,0}_{2,0}\quad,\qquad Y_{4,1;0,0}=F^{1,0}_{3,0}\quad,\qquad Y_{5,1;0,0}=F^{1,0}_{4,0}\quad,\qquad Y_{4,2;0,0}=F^{2,0}_{2,0}\\
Y_{5,2;0,0}&=F^{2,0}_{3,0}\quad,\qquad Y_{5,3;0,0}=F^{3,0}_{2,0}\quad,\qquad Y_{6,4;0,0}=F^{4,0}_{2,0}
\end{align*}
Some results are listed in section~\ref{eigenstates}. From these data, we can fix several coefficients and then conjecture the full collective field Hamiltonian. We will describe it in the next section. Also, we can give the geometrical meaning of these coefficients $A,B,C,D,F,G$ in section~\ref{connectiontomv}, which can also support our conjecture for the collective field Hamiltonian.

\subsection{Hamiltonian}\label{hamiltonian}

\begin{align}
L_0=H_0+H_1+\overline{H}_1+H_2+\overline{H}_2+H_3+\overline{H}_3+H_4+\overline{H_4}+H_5+\overline{H}_5+H_6+\overline{H}_6
\end{align}
\begin{align}
\begin{split}
H_0=&\frac{\lambda}{2}\left(M+\overline{M}\right)-\frac{1}{2N}\left(M-\overline{M}\right)^2-\frac{\lambda}{2N^2}\left(M-\overline{M}+2K-2\overline{K}\right)\left(M-\overline{M}\right)\\
&+\frac{\frac{\lambda^2}{N^2}}{1+\frac{\lambda}{N}}\left[\frac{N}{2}\left(K+\overline{K}\right)-\frac{1}{2N}\left(K-\overline{K}\right)^2\right]
\end{split}
\end{align}
\begin{align}
H_1=\frac{1}{2}M,\qquad \overline{H}_1=\frac{1}{2}\overline{M}
\end{align}
$H_1$, and $\overline{H}_1$ come from $\frac{1}{2}\sum_{i=1}^{N-1}\left(s^+_i-s^-_i\right)^2$ and $\frac{1}{2}\sum_{i=1}^{N-1}\left(r^+_i-r^-_i\right)^2$, respectively. In fact, $H_1$ is not equal to $\frac{1}{2}\sum_{i=1}^{N-1}\left(s^+_i-s^-_i\right)^2$ in general. However, for $\left(S_+;S_-\right)$ and $\left(R_+;R_-\right)$ where $S_+$ and $R_+$ have at most one more box than $S_-$ and $R_-$ at each row, respectively,
\begin{align}
\frac{1}{2}\sum_{i=1}^{N-1}\left(s^+_i-s^-_i\right)^2=\frac{1}{2}M=H_1\;,\qquad\frac{1}{2}\sum_{i=1}^{N-1}\left(r^+_i-r^-_i\right)^2=\frac{1}{2}\overline{M}=\overline{H}_1
\end{align}
$H_0$ and $H_1$ correspond to non-perturbed Hamiltonian, which are composed of global variables, $M, \overline{M}, K, \overline{K}$.

\begin{align}
\begin{split}
H_2=\frac{\frac{\lambda^2}{N^2}}{1+\frac{\lambda}{N}}&\left[ \frac{1}{2}\sum_{n=2}^\infty\sum_{m=1}^{n-1}\sqrt{nm(n-m)}\omega_m\omega_{n-m}\frac{\partial}{\partial \omega_n} +\frac{1}{2}\sum_{n=1}^\infty\sum_{m=1}^{\infty}\sqrt{nm(n+m)}\omega_{n+m}\frac{\partial^2}{\partial \omega_n\partial \omega_m}\right]
\end{split}
\end{align}
\begin{align}
\begin{split}
H_3&=\frac{\lambda}{N}\left[\sum_{n=0}^\infty\sum_{m=1}^\infty \sqrt{m}\psi_{n+m}\frac{\partial^2}{\partial \psi_n\partial \omega_m}+\sum_{n=1}^\infty\sum_{m=1}^{n} \sqrt{m}\psi_{n-m}\omega_m\frac{\partial}{\partial \psi_n}\right]
\end{split}
\end{align}
\begin{align}
\begin{split}
H_4&=\frac{\frac{\lambda^2}{N^2}}{1+\frac{\lambda}{N}}\left[\sum_{n=0}^\infty\sum_{m=1}^\infty n\sqrt{m}\psi_{n+m}\frac{\partial^2}{\partial \psi_n\partial \omega_m}+\sum_{n=1}^\infty\sum_{m=1}^n \left(n-m\right)\sqrt{m}\psi_{n-m}\omega_m\frac{\partial}{\partial \psi_n}\right]
\end{split}
\end{align}
\begin{align}
H_5=-\frac{\lambda}{2N}\sum_{n=0}^\infty\sum_{m=0}^\infty \sum_{u=0}^{n+m}\psi_{n+m-u}\psi_u\frac{\partial^2}{\partial \psi_n\partial \psi_m}
\end{align}
\begin{align}
H_6=-\frac{\frac{\lambda^2}{N^2}}{2\left(1+\frac{\lambda}{N}\right)}\sum_{n=0}^\infty\sum_{m=0}^\infty \sum_{u=0}^{n+m}F\left(n,m,u\right)\psi_{n+m-u}\psi_u\frac{\partial^2}{\partial \psi_n\partial \psi_m}
\end{align}
where
\begin{align}
F\left(n,m,u\right)=\begin{cases}
u &\qquad 0\leqq u\leqq \mbox{min}\left(n,m\right)\\
\mbox{min}\left(n,m\right)&\qquad \mbox{min}\left(n,m\right)\leqq u\leqq \mbox{max}\left(n,m\right)\\
n+m-u&\qquad  \mbox{max}\left(n,m\right)\leqq u \leqq n+m
\end{cases}
\end{align}
$\overline{H}_n$ can be obtained from $H_n$ by substituting $\psi_n$ and $\omega_m$ with $\overline{\psi}_n$ and $\overline{\omega}_m$, respectively. $(n=2,3,4,5,6)$

\section{Eigenstates and Multi-trace Primaries }\label{eigenstates}

According to our methodology, the eigenstates generated by the Hamiltonian will provide exact conformal dimensions and generate all the multi-trace primaries.

In this section, we will analyze these in detail. For a special case, $\left(\Lambda;\Lambda\right)$ (e.g. $\Lambda=\Lambda_+=\Lambda_-$) is expressed in terms of only $\omega$'s. For $\Lambda=\left(\overline{R},S\right)$, a eigenstate $\left(\Lambda;\Lambda\right)$ is just a Schur polynomial.
\begin{align}
\left(\Lambda;\Lambda\right)=P_n\left(S;\left\{\gamma_i\right\}\right)P_m\left(R;\left\{\overline{\gamma}_i\right\}\right)
\end{align}
where 
\begin{align}
n=\left|S\right|,\qquad m=\left|R\right|,\qquad \gamma_j=\sqrt{j}\omega_j
\end{align}
Especially, consider $R_\pm=0$ case.
\begin{align}
\left(S;S\right)=P_n\left(S;\left\{\gamma_i\right\}\right)\equiv\frac{1}{n!}\sum_{g\in S_n}\left[ch_S(g)\prod_{i=1}^\infty\left(\gamma_i\right)^{\lambda(g)_i}\right]\label{lightstate}
\end{align}
where $\left|S\right|=n$ and $ch_\Lambda(g)$ is a character of $g\in S_n$ in the representation of $\Lambda$. A conjugate class of $g\in S_n$ can be expressed as Young tableau. This Young tableau is parametrized by $r_i$, the number of boxes in the $i$th row. Then, define
\begin{align}
\lambda\left(g\right)_i\equiv r_i-r_{i+1}
\end{align}
For example, eigenstates in $Y_{1,1;0;0}$ is 
\begin{align*}
\left(\;{\tiny\yng(1)}\;;\;{\tiny\yng(1)}\;\right)=&\gamma_1
\end{align*}
and a corresponding conformal dimension is
\begin{align*}
h\left(\;{\tiny\yng(1)}\;;\;{\tiny\yng(1)}\;\right)=\epsilon\left(\frac{1}{2}N-\frac{1}{2N}\right)
\end{align*}
where $\epsilon=\frac{\frac{\lambda^2}{N^2}}{1+\frac{\lambda}{N}}$. For $Y_{2,2;0,0}$, we have
\begin{align*}
\left(\;{\tiny\yng(1,1)}\;;\;{\tiny\yng(1,1)}\;\right)=&\frac{1}{2}\left(\gamma_1^2-\gamma_2\right)\\
\left(\;{\tiny\yng(2)}\;;\;{\tiny\yng(2)}\;\right)=&\frac{1}{2}\left(\gamma_1^2+\gamma_2\right)
\end{align*}
\begin{align*}
h\left(\;{\tiny\yng(1,1)}\;;\;{\tiny\yng(1,1)}\;\right)=&\epsilon\left(N-1-\frac{2}{N}\right)\\
h\left(\;{\tiny\yng(2)}\;;\;{\tiny\yng(2)}\;\right)=&\epsilon\left(N+1-\frac{2}{N}\right)
\end{align*}
For $Y_{3,3;0,0}$,
\begin{align*}
\left(\;{\tiny\yng(1,1,1)}\;;\;{\tiny\yng(1,1,1)}\;\right)=&\frac{1}{6}\left(\gamma_1^3-3\gamma_1\gamma_2+2\gamma_3\right)\\
\left(\;{\tiny\yng(2,1)}\;;\;{\tiny\yng(2,1)}\;\right)=&\frac{1}{3}\left(\gamma_1^3-\gamma_3\right)\\
\left(\;{\tiny\yng(3)}\;;\;{\tiny\yng(3)}\;\right)=&\frac{1}{6}\left(\gamma_1^3+3\gamma_1\gamma_2+2\gamma_3\right)
\end{align*}
\begin{align*}
h\left(\;{\tiny\yng(1,1,1)}\;;\;{\tiny\yng(1,1,1)}\;\right)=&\epsilon\left(\frac{3}{2}N-3-\frac{9}{2N}\right)\\
h\left(\;{\tiny\yng(2,1)}\;;\;{\tiny\yng(2,1)}\;\right)=&\epsilon\left(\frac{3}{2}N-\frac{9}{2N}\right)\\
h\left(\;{\tiny\yng(3)}\;;\;{\tiny\yng(3)}\;\right)=&\epsilon\left(\frac{3}{2}N+3-\frac{9}{2N}\right)
\end{align*}

Now, consider general eigenstates. In appendix~\ref{appendixcountingstates}, we claimed that
\begin{align}
F^{k,\overline{k}}_{m,\overline{m}}=Y_{s_+,s_-;r_+,r_-}
\end{align}
where $s_\pm\equiv\left|S_\pm\right|$ and $r_\pm\equiv\left|R_\pm\right|$ with identity
\begin{subequations}
\begin{align}
&m\equiv s_+-s_-\qquad k=s_-\\
&\overline{m}\equiv r_+-r_-\qquad\overline{k}=r_-
\end{align}
\end{subequations}
Especially, they can be interpreted as
\begin{subequations}
\begin{align}
m&= s_+-s_-=(\text{The number of }\psi)\\
k&=s_-=(\text{The total winding number of }\psi, \omega)
\end{align}
\end{subequations}
and, this is similar for $\overline{m}$ and $\overline{k}$. Thus, $\left(\Lambda_+;\Lambda_-\right)$ can be expressed in terms of $\psi,\overline{\psi},\omega$ and $\overline{\omega}$. And, they have the following form
\begin{align}
\begin{split}
\left(\Lambda_+;\Lambda_-\right)=&\left(\sum_{\substack{\{a_i,b_j\}\\\sum_{i=1}^Ma_i+\sum_{j}b_j=K}}c\left(\{a_i,b_j\}\right)\prod_{i=1}^M\psi_{a_{i}}\prod_{j}\omega_{b_j}\right)\left(\sum_{\substack{\{\overline{a}_i,\overline{b}_j\}\\\sum_{i=1}^{\overline{M}}\overline{a}_i+\sum_{j}\overline{b}_j=\overline{K}}}d\left(\{\overline{a}_i,\overline{b}_j\}\right)\prod_{i}^{\overline{M}}\overline{\psi}_{\overline{a}_i}\prod_{j}\overline{\omega}_{\overline{b}_j}\right)
\end{split}
\end{align}
where $c\left(\{a_i,b_j\}\right)$ and $d\left(\{\overline{a}_i,\overline{b}_j\}\right)$ are coefficients. Note that every term has the same winding number. For example,
\begin{align}
\prod_{i=1}^M\psi_{a_i}\prod_{j}\omega_{b_j}\Longrightarrow \sum_{i=1}^M a_i+\sum_{j} b_j=K
\end{align}
Especially, there are some terms in which $\psi$'s do not carry winding number at all. (e.g. $a_i=0$ for $i=1,2,\cdots, M$.) These terms have the following form.
\begin{align}
\psi_0^M\prod_{j}\omega_{b_j},\qquad \text{where}\quad\sum_{j}b_j=K
\end{align}
The coefficients of these terms are related to Schur polynomials.\footnote{In fact, for eigenstates, there is alway ambiguity in choosing overall phase. We determine this overall phase in such a way that the ratio of both side of \eqref{thelastterm} is positive real number.} Considering these terms, we have\footnote{Note that $\omega_n\equiv \frac{1}{\sqrt{n}}\gamma_n$}
\begin{subequations}
\label{thelastterm}
\begin{align}
\sum_{\substack{\{b_j\}\\\sum_jb_j=K}}c\left(\left\{a_i=0,b_j\right\}\right)\psi_0^M \prod_{j}\omega_{b_j}&\sim P_{n=\left|S_-\right|}\left(S_-;\left\{\gamma_i\right\}\right)\\
\sum_{\substack{\{\overline{b}_j\}\\\sum_j\overline{b}_j=\overline{K}}}c\left(\left\{\overline{a}_i=0,\overline{b}_j\right\}\right)\overline{\psi}_0^{\overline{M}} \prod_{j}\overline{\omega}_{\overline{b}_j}&\sim P_{m=\left|R_-\right|}\left(R_-;\left\{\overline{\gamma}_i\right\}\right)
\end{align}
\end{subequations}
where $P_n$ is a Schur polynomial which is defined as
\begin{align}
P_n(\Lambda;\{x_i\})\equiv\frac{1}{n!}\sum_{g\in S_n}\left[ch_\Lambda(g)\prod_{i=1}^\infty\left(x_i\right)^{\lambda(g)_i}\right]
\end{align}
For instance, eigenstates in $Y_{2,1;0,0}$ are
\begin{align*}
(\;{\tiny\yng(1,1)}\;;\;{\tiny\yng(1)}\;)&=\frac{1}{\sqrt{2}}\left(-\psi_1+\psi_0\omega_1\right)\\
(\;{\tiny\yng(2)}\;;\;{\tiny\yng(1)}\;)&=\frac{1}{\sqrt{2}}\left(\psi_1+\psi_0\omega_1\right)
\end{align*}
and, corresponding conformal dimensions are
\begin{align*}
h(\;{\tiny\yng(1,1)}\;;\;{\tiny\yng(1)}\;)&=E_{2,1}-\frac{\lambda}{N}\\
h(\;{\tiny\yng(2)}\;;\;{\tiny\yng(1)}\;)&=E_{2,1}+\frac{\lambda}{N}
\end{align*}
where $E_{2,1}=\frac{1}{2}\left(1+\lambda\right)-\frac{1}{2N}-\frac{3\lambda}{2N^2}+\epsilon\left(\frac{1}{2}N-\frac{1}{2N}\right)$.

In addition, eigenstates in $Y_{3,2;0,0}$ are
\begin{subequations}\label{eigenstateexample1}
\begin{align}
(\;{\tiny\yng(1,1,1)}\;;\;{\tiny\yng(1,1)}\;)&=\frac{1}{\sqrt{3}}\psi_2-\frac{1}{\sqrt{3}}\psi_1\omega_1+\frac{1}{2\sqrt{3}}\psi_0\omega_1^2-\frac{1}{\sqrt{6}}\psi_0\omega_2\\
(\;{\tiny\yng(2,1)}\;;\;{\tiny\yng(2)}\;)&=-\frac{1}{\sqrt{6}}\psi_2-\frac{1}{\sqrt{6}}\psi_1\omega_1+\frac{1}{\sqrt{6}}\psi_0\omega_1^2+\frac{1}{\sqrt{3}}\psi_0\omega_2\\
(\;{\tiny\yng(2,1)}\;;\;{\tiny\yng(1,1)}\;)&=-\frac{1}{\sqrt{6}}\psi_2+\frac{1}{\sqrt{6}}\psi_1\omega_1+\frac{1}{\sqrt{6}}\psi_0\omega_1^2-\frac{1}{\sqrt{3}}\psi_0\omega_2\\
(\;{\tiny\yng(3)}\;;\;{\tiny\yng(2)}\;)&=\frac{1}{\sqrt{3}}\psi_2+\frac{1}{\sqrt{3}}\psi_1\omega_1+\frac{1}{2\sqrt{3}}\psi_0\omega_1^2+\frac{1}{\sqrt{6}}\psi_0\omega_2
\end{align}
\end{subequations}
and corresponding conformal dimensions are
\begin{align*}
h(\;{\tiny\yng(1,1,1)}\;;\;{\tiny\yng(1,1)}\;)&=E_{3,2}-2\frac{\lambda}{N}-\epsilon\\
h(\;{\tiny\yng(2,1)}\;;\;{\tiny\yng(2)}\;)&=E_{3,2}-\frac{\lambda}{N}+\epsilon\\
h(\;{\tiny\yng(2,1)}\;;\;{\tiny\yng(1,1)}\;)&=E_{3,2}+\frac{\lambda}{N}-\epsilon\\
h(\;{\tiny\yng(3)}\;;\;{\tiny\yng(2)}\;)&=E_{3,2}+2\frac{\lambda}{N}+\epsilon
\end{align*}
where $E_{3,2}=\frac{1}{2}\left(1+\lambda\right)-\frac{1}{2N}-\frac{5\lambda}{2N^2}+\epsilon\left(N-\frac{2}{N}\right)$. And, the last two terms in each \eqref{eigenstateexample1} are $\psi_0^2$ times a Schur polynomial of $\omega$'s of degree 2.

For $Y_{4,3;0,0}$,
\begin{align*}
(\;{\tiny\yng(1,1,1,1)}\;;\;{\tiny\yng(1,1,1)}\;)&=-\frac{1}{2}\psi_3+\frac{1}{2}\psi_2\omega_1-\frac{1}{4}\psi_1\omega_1^2+\frac{1}{2\sqrt{2}}\psi_1\omega_2+\frac{1}{12}\psi_0\omega_1^3-\frac{1}{2\sqrt{2}}\psi_0\omega_1\omega_2+\frac{1}{2\sqrt{3}}\psi_0\omega_3\\
(\;{\tiny\yng(2,1,1)}\;;\;{\tiny\yng(2,1)}\;)&=\frac{1}{\sqrt{6}}\psi_3+\frac{1}{2\sqrt{6}}\psi_2\omega_1-\frac{1}{\sqrt{6}}\psi_1\omega_1^2-\frac{1}{2\sqrt{3}}\psi_1\omega_2+\frac{1}{2\sqrt{6}}\psi_0\omega_1^3-\frac{1}{2\sqrt{2}}\psi_0\omega_3\\
(\;{\tiny\yng(3,1)}\;;\;{\tiny\yng(3)}\;)&=-\frac{1}{2\sqrt{3}}\psi_3-\frac{1}{2\sqrt{3}}\psi_2\omega_1-\frac{1}{4\sqrt{3}}\psi_1\omega_1^2-\frac{1}{2\sqrt{6}}\psi_1\omega_2+\frac{1}{4\sqrt{3}}\psi_0\omega_1^3+\frac{3}{2\sqrt{6}}\psi_0\omega_1\omega_2+\frac{1}{2}\psi_0\omega_3\\
(\;{\tiny\yng(2,2)}\;;\;{\tiny\yng(2,1)}\;)&=-\frac{1}{2}\psi_2\omega_1+\frac{1}{\sqrt{2}}\psi_1\omega_2+\frac{1}{6}\psi_0\omega_1^3-\frac{1}{2\sqrt{3}}\psi_0\omega_3\\
(\;{\tiny\yng(2,1,1)}\;;\;{\tiny\yng(1,1,1)}\;)&=\frac{1}{2\sqrt{3}}\psi_3-\frac{1}{2\sqrt{3}}\psi_2\omega_1+\frac{1}{4\sqrt{3}}\psi_1\omega_1^2-\frac{1}{2\sqrt{6}}\psi_1\omega_2+\frac{1}{4\sqrt{3}}\psi_0\omega_1^3-\frac{3}{2\sqrt{6}}\psi_0\omega_1\omega_2+\frac{1}{2}\psi_0\omega_3\\
(\;{\tiny\yng(3,1)}\;;\;{\tiny\yng(2,1)}\;)&=-\frac{1}{\sqrt{6}}\psi_3+\frac{1}{2\sqrt{6}}\psi_2\omega_1+\frac{1}{\sqrt{6}}\psi_1\omega_1^2-\frac{1}{2\sqrt{3}}\psi_1\omega_2+\frac{1}{2\sqrt{6}}\psi_0\omega_1^3-\frac{1}{2\sqrt{2}}\psi_0\omega_3\\
(\;{\tiny\yng(4)}\;;\;{\tiny\yng(3)}\;)&=\frac{1}{2}\psi_3+\frac{1}{2}\psi_2\omega_1+\frac{1}{4}\psi_1\omega_1^2+\frac{1}{2\sqrt{2}}\psi_1\omega_2+\frac{1}{12}\psi_0\omega_1^3+\frac{1}{2\sqrt{2}}\psi_0\omega_1\omega_2+\frac{1}{2\sqrt{3}}\psi_0\omega_3
\end{align*}
\begin{align*}
h(\;{\tiny\yng(1,1,1,1)}\;;\;{\tiny\yng(1,1,1)}\;)&=E_{4,3}-3\frac{\lambda}{N}-3\epsilon\\
h(\;{\tiny\yng(2,1,1)}\;;\;{\tiny\yng(2,1)}\;)&=E_{4,3}-2\frac{\lambda}{N}\\
h(\;{\tiny\yng(3,1)}\;;\;{\tiny\yng(3)}\;)&=E_{4,3}-\frac{\lambda}{N}+3\epsilon\\
h(\;{\tiny\yng(2,2)}\;;\;{\tiny\yng(2,1)}\;)&=E_{4,3}+\epsilon\\
h(\;{\tiny\yng(2,1,1)}\;;\;{\tiny\yng(1,1,1)}\;)&=E_{4,3}+\frac{\lambda}{N}-3\epsilon\\
h(\;{\tiny\yng(3,1)}\;;\;{\tiny\yng(2,1)}\;)&=E_{4,3}+2\frac{\lambda}{N}\\
h(\;{\tiny\yng(4)}\;;\;{\tiny\yng(3)}\;)&=E_{4,3}+3\frac{\lambda}{N}+3\epsilon
\end{align*}
where $E_{4,3}=\frac{1}{2}\left(1+\lambda\right)-\frac{1}{2N}-\frac{7\lambda}{2N^2}+\epsilon\left(\frac{3N}{2}-\frac{9}{2N}\right)$.

For $Y_{4,2;0,0}$,
\begin{align*}
(\;{\tiny\yng(1,1,1,1)}\;;\;{\tiny\yng(1,1)}\;)&=\frac{1}{\sqrt{3}}\psi_2\psi_0+\frac{1}{2\sqrt{3}}\psi_1^2-\frac{1}{\sqrt{3}}\psi_1\psi_0\omega_1+\frac{1}{4\sqrt{3}}\psi_0^2\omega_1^2-\frac{1}{2\sqrt{6}}\psi_0^2\omega_2\\
(\;{\tiny\yng(2,1,1)}\;;\;{\tiny\yng(2)}\;)&=-\frac{1}{2}\psi_2\psi_0-\frac{1}{2}\psi_1\psi_0\omega_1+\frac{1}{4}\psi_0^2\omega_1^2+\frac{1}{2\sqrt{2}}\psi_0^2\omega_2\\
(\;{\tiny\yng(2,1,1)}\;;\;{\tiny\yng(1,1)}\;)&=-\frac{1}{2}\psi_1^2+\frac{1}{4}\psi_0^2\omega_1^2-\frac{1}{2\sqrt{2}}\psi_0^2\omega_2\\
(\;{\tiny\yng(2,2)}\;;\;{\tiny\yng(1,1)}\;)&=-\frac{1}{\sqrt{6}}\psi_2\psi_0+\frac{1}{\sqrt{6}}\psi_1^2+\frac{1}{\sqrt{6}}\psi_1\psi_0\omega_1+\frac{1}{2\sqrt{6}}\psi_0^2\omega_1^2-\frac{1}{2\sqrt{3}}\psi_0^2\omega_2\\
(\;{\tiny\yng(3,1)}\;;\;{\tiny\yng(2)}\;)&=\frac{1}{2}\psi_2\psi_0+\frac{1}{2}\psi_1\psi_0\omega_1+\frac{1}{4}\psi_0^2\omega_1^2+\frac{1}{2\sqrt{2}}\psi_0^2\omega_2\\
\end{align*}
\begin{align*}
h(\;{\tiny\yng(1,1,1,1)}\;;\;{\tiny\yng(1,1)}\;)&=E_{4,2}-5\frac{\lambda}{N}-\epsilon\\
h(\;{\tiny\yng(2,1,1)}\;;\;{\tiny\yng(2)}\;)&=E_{4,2}-3\frac{\lambda}{N}+\epsilon\\
h(\;{\tiny\yng(2,1,1)}\;;\;{\tiny\yng(1,1)}\;)&=E_{4,2}-\frac{\lambda}{N}-\epsilon\\
h(\;{\tiny\yng(2,2)}\;;\;{\tiny\yng(1,1)}\;)&=E_{4,2}+\frac{\lambda}{N}-\epsilon\\
h(\;{\tiny\yng(3,1)}\;;\;{\tiny\yng(2)}\;)&=E_{4,2}+\frac{\lambda}{N}+\epsilon
\end{align*}
where $E_{4,2}=\frac{3}{2}\left(1+\lambda\right)-\frac{9}{2N}-\frac{21\lambda}{2N^2}+\epsilon\left(N-\frac{2}{N}\right)$.

For $Y_{3,2;0,0}$,
\begin{align*}
(\;{\tiny\yng(1,1,1,1,1)}\;;\;{\tiny\yng(1,1)}\;)&=\frac{3}{2\sqrt{15}}\psi_2\psi_0^2+\frac{3}{2\sqrt{15}}\psi_1^2\psi_0-\frac{3}{2\sqrt{15}}\psi_1\psi_0^2\omega_1+\frac{1}{4\sqrt{15}}\psi_0^3\omega_1^2-\frac{1}{2\sqrt{30}}\psi_0^3\omega_2\\
(\;{\tiny\yng(2,1,1,1)}\;;\;{\tiny\yng(2)}\;)&=-\frac{3}{2\sqrt{15}}\psi_2\psi_0^2-\frac{3}{2\sqrt{15}}\psi_1\psi_0^2\omega_1+\frac{1}{2\sqrt{15}}\psi_0^3\omega_1^2+\frac{1}{\sqrt{30}}\psi_0^3\omega_2\\
(\;{\tiny\yng(2,1,1,1)}\;;\;{\tiny\yng(1,1)}\;)&=\frac{1}{2\sqrt{15}}\psi_2\psi_0^2-\frac{2}{\sqrt{15}}\psi_1^2\psi_0-\frac{1}{2\sqrt{15}}\psi_1\psi_0^2\omega_1+\frac{1}{2\sqrt{15}}\psi_0^3\omega_1^2-\frac{1}{\sqrt{30}}\psi_0^3\omega_2\\
(\;{\tiny\yng(2,2,1)}\;;\;{\tiny\yng(1,1)}\;)&=-\frac{1}{2\sqrt{3}}\psi_2\psi_0^2+\frac{1}{2\sqrt{3}}\psi_1^2\psi_0+\frac{1}{2\sqrt{3}}\psi_1\psi_0^2\omega_1+\frac{1}{4\sqrt{3}}\psi_0^3\omega_1^2-\frac{1}{2\sqrt{6}}\psi_0^3\omega_2\\
(\;{\tiny\yng(3,1,1)}\;;\;{\tiny\yng(2)}\;)&=\frac{1}{\sqrt{10}}\psi_2\psi_0^2+\frac{1}{\sqrt{10}}\psi_1\psi_0^2\omega_1+\frac{1}{2\sqrt{10}}\psi_0^3\omega_1^2+\frac{1}{2\sqrt{5}}\psi_0^3\omega_2\\
\end{align*}
\begin{align*}
h(\;{\tiny\yng(1,1,1,1,1)}\;;\;{\tiny\yng(1,1)}\;)&=E_{5,2}-9\frac{\lambda}{N}-\epsilon\\
h(\;{\tiny\yng(2,1,1,1)}\;;\;{\tiny\yng(2)}\;)&=E_{5,2}-6\frac{\lambda}{N}+\epsilon\\
h(\;{\tiny\yng(2,1,1,1)}\;;\;{\tiny\yng(1,1)}\;)&=E_{5,2}-4\frac{\lambda}{N}-\epsilon\\
h(\;{\tiny\yng(2,2,1)}\;;\;{\tiny\yng(1,1)}\;)&=E_{5,2}-\frac{\lambda}{N}-\epsilon\\
h(\;{\tiny\yng(3,1,1)}\;;\;{\tiny\yng(2)}\;)&=E_{5,2}-\frac{\lambda}{N}+\epsilon
\end{align*}
where $E_{5,2}=\frac{3}{2}\left(1+\lambda\right)-\frac{9}{2N}-\frac{21\lambda}{2N^2}+\epsilon\left(N-\frac{2}{N}\right)$.

\section{ Matrix-vector Model and Geometric Picture}\label{connectiontomv}

This Hamiltonian has several central features. First of all, it operates in a Fock space with one extra dimension represented by the winding number coordinate $n$. It was shown in the previous section that it reproduces the nonlinear primaries as exact eigenstates with exact eigenvalues. Furthermore, there is an exact, relatively surprising correspondence with matrix-vector models. This will imply that the theory exhibits locality in terms of the coordinate $\zeta$ conjugate to winding number as in the case of matrix models \cite{Das:1990kaa}.

Let us begin by discussing in more detail the correspondence with the matrix-vector model interactions and their geometric interpretation.
In the full Hamiltonian, there are two types of interaction coupling constants, $\epsilon=\frac{\frac{\lambda^2}{N^2}}{1+\frac{\lambda}{N}}$ and $g=\frac{\lambda}{N}$. $H_2, H_4$ and $H_6$ are proportional to $\epsilon$ while $H_3$ and $H_5$ have $g$ as coupling constant. e.g. $H=H_{\text{global}}+H_{\epsilon}+H_g$ where $H_{\text{global}}=2H_0$, $H_\epsilon=2H_2+2H_4+2H_6$ and $H_g=2H_3+2H_5$. In addition, $H_2, H_4$ and $H_6$ have different properties from $H_3$ and $H_5$. 

$H_3$ and $H_5$ are interactions of $\psi_m$ $(m=0,1,2,\cdots)$ and $\omega_n$ $(n=1,2,\cdots)$. On the other hands, $H_2, H_4$ and $H_6$ are independent of $\psi_0$. For example, coefficients related to $\psi_0$ are zero in $H_2, H_4$ and $H_6$. In fact, $H_3$ and $H_5$ are related to extra terms when we shift indices of $\psi_n$ in $H_4$ and $H_6$ by $-1$. In detail, Under shift $\psi_n\longrightarrow \psi_{n-1}$,
\begin{align}
\begin{split}
&A\equiv\sum_{n,m=1}^\infty nm\psi_{n+m}\frac{\partial^2}{\partial \gamma_n\partial \psi_m}+\sum_{n=1}^\infty\sum_{m=1}^n\left(n-m\right)\gamma_m\psi_{n-m}\frac{\partial}{\partial \psi_n}\\
&\longrightarrow\; A+\sum_{m=0}^\infty\sum_{n=1}^\infty n\psi_{n+m}\frac{\partial^2}{\partial \gamma_{n}\partial \psi_{m}}+\sum_{n=1}^\infty\sum_{m=1}^{n}\gamma_{m}\psi_{n-m}\frac{\partial}{\partial \psi_{n}}
\end{split}
\end{align}
\begin{align}
\begin{split}
&B\equiv\frac{1}{2}\sum_{n,m=0}^\infty\sum_{u=0}^{n+m}F\left(n,m,u\right)\psi_{n+m-u}\psi_u\frac{\partial^2}{\partial\psi_n\partial\psi_m}\;\longrightarrow\; B+\sum_{n,m=0}^\infty\sum_{u=0}^{n+m} \psi_{n+m-u }\psi_{u}\frac{\partial^2}{\partial\psi_{n}\partial\psi_{m}}
\end{split}
\end{align}
Later, we can see that $H_2, H_4$ and $H_6$ have good geometrical interpretation such as joining and splitting of loops, whereas $H_3$ and $H_5$ correspond to special boundary terms.

The theory is expandable in the different limits. In the t'Hooft limit, one has
\begin{align}
c=\left(N-1\right)\left[1-\frac{N\left(N+1\right)}{\left(N+k\right)\left(N+k+1\right)}\right]
\end{align}
and $\lambda=\frac{N}{N+k}$ is fixed while $N$ is taken to be large. Consequently,
\begin{align}
g=\frac{\lambda}{N}\gg \frac{\frac{\lambda^2}{N^2}}{1+\frac{\lambda}{N}}=\epsilon
\end{align}
Hence, $H_3$ and $H_5$ are larger than $H_2, H_4$ and $H_6$ in the t'Hooft limit. On the other hands, \cite{gaberdielgopakumar2012} and \cite{perlmutter2012} proposed the semiclassical limit in which central charge $c$ is taken to be large with finite $N$. The coupling constants become
\begin{align}
\frac{\lambda}{N}=-1+\mathcal{O}\left(\frac{1}{c}\right)\;,\qquad \frac{\frac{\lambda^2}{N^2}}{1+\frac{\lambda}{N}}=-\frac{c}{N\left(N^2-1\right)}+\mathcal{O}\left(1\right)
\end{align}
Thus,  
\begin{align}
-\frac{\frac{\lambda^2}{N^2}}{1+\frac{\lambda}{N}}\sim \frac{c}{N\left(N^2-1\right)}\gg\frac{\lambda}{N}\sim \mathcal{O}\left(1\right)
\end{align}
Therefore, $H_2, H_4$ and $H_6$ are dominant terms in the semiclassical limit up to global variables. 

It is not coincident that $H_2, H_4$ and $H_6$ have such good properties. They are equal to the Hamiltonian of the matrix-vector model in \cite{antal1996}. \cite{antal1996} considered $SU\left(N\right)$ matrix and complex vector fields with a Hamiltonian,
\begin{align}
H_{\text{MV}}=\frac{1}{2}m_M\sum_{i,j=1}^N \frac{\partial^2}{\partial U_{ij}\partial U_{ji}}+\frac{1}{2}\sum_{a=1}^{n_f}\sum_{i=1}^n m_a\frac{\partial}{\partial \overline{x}^i_a}\frac{\partial}{\partial x^i_a}+\text{potential}
\end{align}
where $a=1,2,\cdots, n_f$ is flavor and $i,j=1,2,\cdots, N$ is color. Under transformation
\begin{align}
U,\; x^a\quad\longrightarrow\quad VUV^{-1},\; Vx^a\qquad\qquad \text{for}\quad V\in U\left(N\right)
\end{align}
one can define invariant collective variables
\begin{align}
\gamma_n\equiv tr\left(U^n\right),\qquad \psi^{ab}_n\equiv \overline{x}^a\cdot U\cdot x^b\label{matrixvectorvariables2}
\end{align}
Then, the Hamiltonian can be expressed in terms of these invariant collective variables.
\begin{align}
\begin{split}
H_{\text{MV}}=&m_N\left(\sum_{n=1}^\infty Nn\gamma_n\frac{\partial}{\partial \gamma_n}+\sum_{n=0}^\infty Nn\psi^{ab}_n\frac{\partial}{\partial \psi^{ab}_n}\right)\\
&+m_N\left(\frac{1}{2}\sum_{n,m=1}^\infty nm\gamma_{n+m}\frac{\partial^2}{\partial \gamma_n\gamma_m}+\frac{1}{2}\sum_{n=2}^\infty\sum_{m=1}^{n-1} \gamma_{n-m}\gamma_{m}\frac{\partial}{\partial \gamma_n}\right)\\
&+m_N\left(\sum_{n,m=1}^\infty\sum_{a,b=1}^M nm\psi^{ab}_{n+m}\frac{\partial^2}{\partial \gamma_n\partial \psi^{ab}_m}+\sum_{n=1}^\infty\sum_{m=1}^n\left(n-m\right)\gamma_m\psi^{ab}_{n-m}\frac{\partial}{\partial \psi^{ab}_n}\right)\\
&+\frac{m_N}{2}\sum_{n,m=0}^\infty\sum_{u=0}^{n+m}\sum_{a,b,c,d=1}^M F\left(n,m,u\right)\psi^{ad}_{n+m-u}\psi^{cb}_u\frac{\partial^2}{\partial\psi^{ab}_n\partial\psi^{cd}_m}+\cdots
\end{split}
\end{align}
where dots mean contributions from Hamiltonian of vector fields and potentials. Take $m_M=-\frac{\frac{\lambda^2}{N^2}}{1+\frac{\lambda}{N}}$(massive matrix field\footnote{Note that $-\frac{\frac{\lambda^2}{N^2}}{1+\frac{\lambda}{N}}$ is positive infinite in the semiclassical limit.}), $m_a=0$ (massless vector field) and $n_f=1$ (one flavor). Moreover, by inserting $\psi_n=-\sqrt{m}\omega_n$, we have
\begin{align}
H_{\text{MV}}=H_2+H_4+H_6-\frac{\frac{\lambda^2}{N^2}}{1+\frac{\lambda}{N}}NK
\end{align}

On the other hands, if we take the semiclassical limit, the dominant terms (of order $\mathcal{O}\left(c\right)$) in our full Hamiltonian is
\begin{align}
L_0\simeq \frac{\frac{\lambda^2}{N^2}}{1+\frac{\lambda}{N}}\left(\frac{N}{2}K-\frac{1}{2N}K^2\right)+H_2+H_4+H_6
\end{align}
where we ignore conjugate fields for simplicity. Thus, we can conclude
\begin{align}
L_0\sim H_{\text{MV}}+\mathcal{O}\left(1\right) \qquad\mbox{up to global variables}
\end{align}

Now, we will give geometrical interpretation of $H_i$. ($i=1,2,\cdots,6$). For this purpose, it is convenient to express $H$'s in terms of $\gamma_n=\sqrt{n}\omega_n$. Based on the connection with matrix-vector model, recall the \eqref{matrixvectorvariables} or \eqref{matrixvectorvariables2}. We can interpreter $\gamma_n$ as a closed loop with $n$ winding number and $\psi_m$ as an open loop with $m$ winding number. In addition, the open loop corresponding to $\psi_m$ has distinguishable two ends because the vector field is complex.

First of all, $H_2$ is an interaction between closed loops. The first term of $H_2$ corresponds to splitting of one closed loop into two closed loops.
\begin{align}
\sum_{n=2}^\infty\sum_{m=1}^{n-1} \frac{1}{2}n\gamma_m\gamma_{n-m}\frac{\partial}{\partial \gamma_n}
\end{align}
The coefficient of the interaction is the number of ways of splitting. On the other hands, the second term of $H_2$ is related to joining of two closed loops into one closed loop.
\begin{align}
\sum_{n=1}^\infty\sum_{m=1}^{\infty}\frac{1}{2}nm\gamma_{n+m}\frac{\partial^2}{\partial \gamma_n\partial \gamma_m}\end{align}

Next, $H_3$ and $H_4$ are different interactions between a closed loop and an open loop. The first terms of $H_3$ and $H_4$ are a joining of a closed loop and an open loop into a new open loop. And, the second terms of $H_3$ and $H_4$ are splitting a open loop into a new open loop and a closed loop. The only difference is the coefficients of interactions, which implies that the method of joining and splitting is different. When we split an open loop, $H_3$ can cut the open loop at one fixed end of the loop while $H_4$ can cut it at the any point except for one fixed end.

\begin{figure}[tbp]
\centering
\includegraphics[width=6cm]{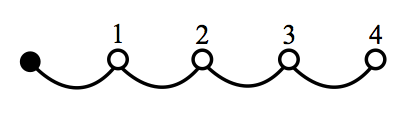}
\centering
\includegraphics[width=3cm]{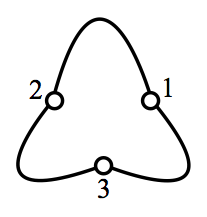}
\caption{The left and right figure are an open loop with winding number 4 and  a closed loop with winding number 3, respectively. For $H_3$, we can cut the open loop only at point 4. In addition, the closed loop can be cut at points 1, 2 and 3. On the other hands, for $H_4$, the open loop can be cut at point 1, 2, 3 and 4. Moreover, we can cut the closed loop at point 1, 2 and 3.}
\label{openclosedloop}
\end{figure}

For instance, consider joining $\psi_4$ and $\gamma_3$. See figure~\ref{openclosedloop}. In the case of $H_3$, we can cut $\psi_4$ only at the point 4 while $\gamma_3$ can be cut at point 1, 2 and 3. Thus, there are three ways to cut. After cutting loops, we can attach each piece to make $\psi_7$. Hence, the number of way to make $\psi_7$ is $3$, which agrees with the coefficient of the joining interaction.

On the other hands, for the joining interaction in $H_4$, $\psi_4$ can be cut at points $1, 2, 3$ and $4$. Moreover, $\gamma_3$ can be cut at points $1, 2$ and $3$. Therefore, the total number of ways to get $\psi_7$ is 12. And, this is the coefficient of the interaction from $\psi_4$ and $\gamma_3$ into $\psi_7$.

Finally, $H_5$ and $H_6$ are different interactions between open loops. $H_5$ corresponds to joining two open loops into other two open loops. However, $H_5$ has less natural interpretation than $H_6$. The coefficient is related to the number of ways of this interaction. That is, the coefficient corresponds to the number of ways to make two open loops with winding number $n+m-u$ and $u$ from the open loops with $n,m$ by ignoring how to cut and attach. However, $H_6$ has more natural geometrical interpretation like $H_2$ or $H_4$. We may suppose that an open loop has distinct two ends denoted by $A, B$ because of complex vector field. You can cut $\psi_n, \psi_m$ at one point except for the one end A, respectively. Then, we have two loops with end $A$ and two loops with end $B$. By attaching two different types of loop, we can get two open loops. 

\begin{figure}[tbp]
\centering
\includegraphics[width=6cm]{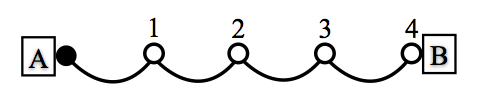}
\centering
\includegraphics[width=3.5cm]{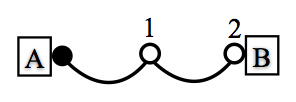}
\caption{Two figure are open loops with winding number 4 and 2, respectively. They have two distinct ends, $A$ and $B$. For $H_6$, we can cut these open loops at the white points. e.g. points 1, 2, 3 and 4 for the left open loop.}
\label{openopenloop}
\end{figure}

For example, consider an example of the interaction from $\psi_4$ and $\psi_2$ into $\psi_1$ and $\psi_5$. See figure~\ref{openopenloop}. $\psi_4$ and $\psi_2$ have two distinguishable ends. Moreover, we can cut open loops once at the white points. e.g. points 1, 2, 3 and 4 for $\psi_4$. After cutting two loops, we can attach each piece to make $\psi_1$ and $\psi_5$. But, a piece with the end $A$ can only be connected to a piece with the end $B$ and vice versa. The total number of possible ways is 2. This is exactly same as $2F\left(4,2,1\right)=2F\left(4,2,5\right)$.

 As in the matrix-vector models, the theory which we constructed exhibits locality in terms of  the coordinate  conjugate to the winding number. One introduces
\begin{align}
\begin{split}
&\gamma\left(\zeta\right)=\sum_{n} \zeta^{-n-1}\gamma_n\;,\quad\Xi\left(z\right)=\sum_{n} \zeta^{n}\frac{\partial}{\partial\gamma_n}\\
&\Psi\left(\zeta\right)=\sum_{n}\zeta^{-n-1}\psi_n\;,\quad\Pi\left(\zeta\right)=\sum_{n}\zeta^n\frac{\partial}{\partial\psi_n}
\end{split}
\end{align}\label{conformalfield}
and, we have
\begin{align}
\alpha_\pm\left(\zeta\right)=\gamma\left(\zeta\right)\pm\partial_\zeta \Xi\left(\zeta\right)
\end{align}
which satisfy the Poisson bracket
\begin{align}
\left\{\alpha_\pm\left(\zeta\right),\alpha_\pm\left(\zeta'\right)\right\}=\pm 2 \partial_\zeta \delta\left(\zeta-\zeta'\right)
\end{align}
The Hamiltonian of matrix-vector model can be expressed in terms of these collective fields. For instance, the cubic interaction is
\begin{align}
H_{\text{cubic}}=\int d\zeta\; \zeta^2\left(\frac{1}{6}\left(\alpha\left(\zeta\right)\right)^3+\alpha\left(\zeta\right)T\left(\zeta\right)\right)
\end{align}
where $T\left(z\right)$ is the energy momentum tensor of the matter fields $\psi$. Altogether our fields are therefore described by the $AdS_3$ space-time and the extra $S_1$ corresponding to the winding coordinate $n$.

\section{Extended Hamiltonian}

The basic construction that we have given is characterized by several main features.  The field theory is constructed to reproduce the nonlinear structure contained in multi-trace primaries with $G=\sfrac{1}{N}$ serving as a coupling constant. It is built on a Fock space associated with single-trace primaries and contains an extra dimension ( the `winding' number ) labeling them. The interactions and vertices ( cubic + quartic ) turned out to be identical in structure to vertices of the large $N$ matrix-vector model \cite{Avan:1995sp}.
It follows then that the field theory is local , when written in terms of the conjugate coordinates in complete parallel with the emergent extra dimensions in $c=1$ models. For the present structure of matrix-vector theories, this locality was established in detail in  \cite{Avan:1996vi} together with an interesting Yangian CFT structure.
So far, we consider states where $\Lambda_-$ is a sub-Young tableau of $\Lambda_+$. In general, we can categorize states $\left(\Lambda_+;\Lambda_-\right)$ into three categories.
\begin{enumerate}
\item $\Lambda_-$ is a sub-Young tableau of $\Lambda_+$.

\item $\Lambda_+$ is a sub-Young tableau of $\Lambda_-$.

\item Neither of them.
\end{enumerate}
Examples of the third category are
\begin{align}
\left(\;{\tiny\yng(2)}\;;\;{\tiny\yng(1,1)}\;\right)\;,\quad \left(\;{\tiny\yng(3)}\;;\;{\tiny\yng(1,1)}\;\right)\;,\quad \left(\;{\tiny\yng(2,1)}\;;\;{\tiny\yng(1,1,1,1)}\;\right)\;,\;\cdots
\end{align}
As we have mentioned before these operators contain derivatives and are not involved in the present study. On the other hands, the second category seems to be parallel to the first category. Indeed we will now exhibit  a symmetry which will allow us to carry over our previous construction to the following states;
\begin{enumerate}
\item $\Lambda_-$ is a sub-Young tableau of $\Lambda_+$. Moreover, in each row, $\Lambda_+$ has at most one more box than $\Lambda_-$. 

\item $\Lambda_+$ is a sub-Young tableau of $\Lambda_-$. In addition, in each row, $\Lambda_-$ has at most one more box than $\Lambda_+$. 
\end{enumerate}
Moreover, a single-trace operator $\widetilde{\psi}$ in the second category corresponds to $\psi$ in the first one while $\omega$ is common in both category.

However, even though $\psi_n$ and $\widetilde{\psi}_m$ look parallel, there is a difference between them. Considering the conformal dimension, we want to keep the definition of total winding number which is the number of boxes in $\Lambda_-$. Hence, we will identify 
\begin{align}
\left(\;0\;;\;{\tiny\yng(1)}\;\right)=\widetilde{\psi}_1
\end{align}
Contrast to $\psi_n$, the index of $\widetilde{\psi}_n$ starts from $1$. Thus, we can not directly use the previous result of $\psi$'s and $\omega$'s, but we must establish again a collective Hamiltonian of $\widetilde{\psi}_n$ and $\omega_n$ in the exactly same way as before. We will omit the procedure to obtain the collective Hamiltonian because it is exactly same way. 

In spite of this slight asymmetry between $\psi$ and $\widetilde{\psi}$, we obtain a collective Hamiltonian of $\widetilde{\psi}$ and $\omega$ which is almost same as the collective Hamiltonian of $\psi$ and $\omega$. In addition, we found symmetry in the eigenstates of both cases. In the next section, we will describe the result.

\subsection{Extension}\label{tildehamiltonian}

Before describing the result, we will rephrase winding number and the number of $\psi$ and $\widetilde{\psi}$. For given Young tableau $\Lambda_-$, we can make $\Lambda_+$ in the following two ways.
\begin{enumerate}
\item In each row of $\Lambda_-$, we can add at most one more box to $\Lambda_-$.

\item In each row of $\Lambda_-$, we can remove at most one more box from $\Lambda_-$.
\end{enumerate}
The first one corresponds to $\psi$, the second one corresponds to $\widetilde{\psi}$. And, if we add or remove no boxes, then it corresponds to $\omega$. However, we can not mix two ways at this stage. For example, we will ignore a possibility adding one box in the first row and removing one box in the second row.

Then, for $\left|\Lambda_+\right|-\left|\Lambda_-\right|>0$,
\begin{align}
\begin{split}
\left|\Lambda_+\right|-\left|\Lambda_-\right|=&\left(\mbox{The number of boxes added to $\Lambda_-$ to make $\Lambda_+$}\right)\\
=&\left(\mbox{The number of $\psi$'s}\right)
\end{split}
\end{align}
On the other hands, for $\left|\Lambda_+\right|-\left|\Lambda_-\right|<0$,
\begin{align}
\begin{split}
\left|\Lambda_-\right|-\left|\Lambda_+\right|=&\left(\mbox{The number of boxes removed from $\Lambda_-$ to make $\Lambda_+$}\right)\\
=&\left(\mbox{The number of $\widetilde{\psi}$'s}\right)
\end{split}
\end{align}
and
\begin{align}
\left|\Lambda_-\right|=\mbox{Total winding number}
\end{align}

The Hamiltonian is
\begin{align}
L_0=H_0+H_1+\overline{H}_1+\widetilde{H}_1+\overline{\widetilde{H}}_1+H_2+\overline{H}_2+\sum_{i=3}^6\left(H_i+\overline{H}_i+\widetilde{H}_i+\overline{\widetilde{H}}_i\right)
\end{align}
\begin{align}
\begin{split}
H_0=&\frac{\lambda}{2}\left(M+\overline{M}-\widetilde{M}-\overline{\widetilde{M}}\right)-\frac{1}{2N}\left(M-\overline{M}\right)^2-\frac{1}{2N}\left(\widetilde{M}-\overline{\widetilde{M}}\right)^2\\
&-\frac{\lambda}{2N^2}\left(M-\overline{M}+2K-2\overline{K}\right)\left(M-\overline{M}\right)-\frac{\lambda}{2N^2}\left(\widetilde{M}-\overline{\widetilde{M}}-2K+2\overline{K}\right)\left(\widetilde{M}-\overline{\widetilde{M}}\right)\\
&+\frac{\frac{\lambda^2}{N^2}}{1+\frac{\lambda}{N}}\left[\frac{N}{2}\left(K+\overline{K}\right)-\frac{1}{2N}\left(K-\overline{K}\right)^2\right]
\end{split}
\end{align}
where
\begin{align}
\widetilde{M}\equiv&\sum_{n=1}^\infty \widetilde{\psi}_n\frac{\partial}{\partial\widetilde{\psi}_n}\;,\qquad\overline{\widetilde{M}}\equiv\sum_{n=1}^\infty \overline{\widetilde{\psi}}_n\frac{\partial}{\partial\overline{\widetilde{\psi}}_n}
\end{align}
and
\begin{subequations}
\begin{align}
K=\sum_{n=0}^\infty n\psi\frac{\partial}{\partial \psi_n}+\sum_{n=1}^\infty n\widetilde{\psi}_n\frac{\partial}{\partial\widetilde{\psi}_n}+\sum_{n=1}^\infty n \omega_i\frac{\partial}{\partial \omega_n}\\
\overline{K}=\sum_{n=0}^\infty n\overline{\psi}\frac{\partial}{\partial \overline{\psi}_n}+\sum_{n=1}^\infty n\overline{\widetilde{\psi}}_n\frac{\partial}{\partial\overline{\widetilde{\psi}}_n}+\sum_{i=1}^\infty i \overline{\omega}_i\frac{\partial}{\partial \overline{\omega}_i}
\end{align}
\end{subequations}
$H_0$ belongs to non-perturbed Hamiltonian which is composed of global variables $M$, $\overline{M}$, $\widetilde{M}$, $\overline{\widetilde{M}}$, $K$ and $\overline{K}$. Moreover, the non-perturbed Hamiltonian also contains
\begin{align}
\widetilde{H_1}=\frac{1}{2}\widetilde{M}
\end{align}
We can express $\widetilde{H}_1$ in terms of global variables only when we impose the condition that we subtract at most one box from $\Lambda_-$. Otherwise, it will have additional contribution related to derivatives.
\begin{align}
\begin{split}
\widetilde{H}_3&=-\frac{\lambda}{N}\left[\sum_{n=1}^\infty\sum_{m=1}^\infty \sqrt{m}\widetilde{\psi}_{n+m}\frac{\partial^2}{\partial \widetilde{\psi}_n\partial \omega_m}+\sum_{n=2}^\infty\sum_{m=1}^{n-1} \sqrt{m}\widetilde{\psi}_{n-m}\omega_m\frac{\partial}{\partial \widetilde{\psi}_n}\right]
\end{split}
\end{align}
\begin{align}
\begin{split}
\widetilde{H}_4&=\frac{\frac{\lambda^2}{N^2}}{1+\frac{\lambda}{N}}\left[\sum_{n=1}^\infty\sum_{m=1}^\infty {n}\sqrt{m}\widetilde{\psi}_{n+m}\frac{\partial^2}{\partial \widetilde{\psi}_n\partial \omega_m}+\sum_{n=2}^\infty\sum_{m=1}^{n-1} \left(n-m\right)\sqrt{m}\widetilde{\psi}_{n-m}\omega_m\frac{\partial}{\partial \widetilde{\psi}_n}\right]
\end{split}
\end{align}
\begin{align}
\widetilde{H}_5=\frac{\lambda}{2N}\sum_{n=1}^\infty\sum_{m=1}^\infty \sum_{u=1}^{n+m-1}\widetilde{\psi}_{n+m-u}\widetilde{\psi}_u\frac{\partial^2}{\partial \widetilde{\psi}_n\partial \widetilde{\psi}_m}
\end{align}
\begin{align}
\widetilde{H}_6=-\frac{\frac{\lambda^2}{N^2}}{2\left(1+\frac{\lambda}{N}\right)}\sum_{n=1}^\infty\sum_{m=1}^\infty \sum_{u=1}^{n+m-1}F\left(n,m,u\right)\widetilde{\psi}_{n+m-u}\widetilde{\psi}_u\frac{\partial^2}{\partial \widetilde{\psi}_n\partial \widetilde{\psi}_m}
\end{align}
where
\begin{align}
F\left(n,m,u\right)=\begin{cases}
u &\qquad 0\leqq u\leqq \mbox{min}\left(n,m\right)\\
\mbox{min}\left(n,m\right)&\qquad \mbox{min}\left(n,m\right)\leqq u\leqq \mbox{max}\left(n,m\right)\\
n+m-u&\qquad  \mbox{max}\left(n,m\right)\leqq u \leqq n+m
\end{cases}
\end{align}
Especially, we can find a connection between $H_i$ and $\widetilde{H}_i$. ($i=3,4,5,6$)
\begin{align}
\begin{split}
\widetilde{H}_3=&-\left.H_3\right|_{\psi_n\rightarrow \widetilde{\psi}_{n+1}}\\
\widetilde{H}_4=&\left.H_4\right|_{\psi_n\rightarrow \widetilde{\psi}_{n+1}}\\
\widetilde{H}_5=&-\left.H_5\right|_{\psi_n\rightarrow \widetilde{\psi}_{n+1}}\\
\widetilde{H}_6=&\left.H_6\right|_{\psi_n\rightarrow \widetilde{\psi}_{n+1}}
\end{split}
\end{align}

\subsection{Extended Eigenstate}

For $\left|\Lambda_+\right|>\left|\Lambda_-\right|$, the eigenstate of $\left(\Lambda_-;\Lambda_+\right)$ is equal to one obtained by replacing $\psi_n$  in $\left(\Lambda_+;\Lambda_-\right)$ with $\widetilde{\psi}_{n+1}$.
\begin{align}
\left(\Lambda_+;\Lambda_-\right)\xrightarrow{\quad\psi_n\;\Longrightarrow \;\widetilde{\psi}_{n+1}\quad} \left(\Lambda_-;\Lambda_+\right)
\end{align}
And, the order of energy is reversed assuming that $\frac{\lambda}{N}>\frac{\frac{\lambda^2}{N^2}}{1+\frac{\lambda}{N}}$. For example,
\begin{alignat*}{5}
&E &&\left(\Lambda_+;\Lambda_-\right) &&\xrightarrow{\quad\psi_n\;\Longrightarrow \;\widetilde{\psi}_{n+1}\quad}&&\left(\Lambda_-;\Lambda_+\right)\qquad &&E\\\\
&E_{3,2}-2\frac{\lambda}{N}-\epsilon\qquad&&\left(\;{\tiny\yng(1,1,1)}\;;\;{\tiny\yng(1,1)}\;\right)&&\xrightarrow{\quad\psi_n\;\Longrightarrow \;\widetilde{\psi}_{n+1}\quad}&&\left(\;{\tiny\yng(1,1)}\;;\;{\tiny\yng(1,1,1)}\;\right)\qquad&&E_{2,3}+2\frac{\lambda}{N}-3\epsilon\\
&E_{3,2}-\frac{\lambda}{N}+\epsilon\qquad&&\left(\;{\tiny\yng(2,1)}\;;\;{\tiny\yng(2)}\;\right)&&\xrightarrow{\quad\psi_n\;\Longrightarrow \;\widetilde{\psi}_{n+1}\quad}&&\left(\;{\tiny\yng(2)}\;;\;{\tiny\yng(2,1)}\;\right)\qquad&&E_{2,3}+\frac{\lambda}{N}\\
&E_{3,2}+\frac{\lambda}{N}-\epsilon\qquad&&\left(\;{\tiny\yng(2,1)}\;;\;{\tiny\yng(1,1)}\;\right)&&\xrightarrow{\quad\psi_n\;\Longrightarrow \;\widetilde{\psi}_{n+1}\quad}&&\left(\;{\tiny\yng(1,1)}\;;\;{\tiny\yng(2,1)}\;\right)\qquad&&E_{2,3}-\frac{\lambda}{N}\\
&E_{3,2}+2\frac{\lambda}{N}+\epsilon\qquad&&\left(\;{\tiny\yng(3)}\;;\;{\tiny\yng(2)}\;\right)&&\xrightarrow{\quad\psi_n\;\Longrightarrow \;\widetilde{\psi}_{n+1}\quad}&&\left(\;{\tiny\yng(2)}\;;\;{\tiny\yng(3)}\;\right)\qquad&&E_{2,3}-2\frac{\lambda}{N}+3\epsilon
\end{alignat*}
\begin{alignat*}{5}
&E &&\left(\Lambda_+;\Lambda_-\right) &&\xrightarrow{\quad\psi_n\;\Longrightarrow \;\widetilde{\psi}_{n+1}\quad}&&\left(\Lambda_-;\Lambda_+\right)\qquad &&E\\\\
&E_{4,2}-5\frac{\lambda}{N}-\epsilon\qquad&&\left(\;{\tiny\yng(1,1,1,1)}\;;\;{\tiny\yng(1,1)}\;\right)&&\xrightarrow{\quad\psi_n\;\Longrightarrow \;\widetilde{\psi}_{n+1}\quad}&&\left(\;{\tiny\yng(1,1)}\;;\;{\tiny\yng(1,1,1,1)}\;\right)\qquad&&E_{2,4}+5\frac{\lambda}{N}-6\epsilon\\
&E_{4,2}-3\frac{\lambda}{N}+\epsilon\qquad&&\left(\;{\tiny\yng(2,1,1)}\;;\;{\tiny\yng(2)}\;\right)&&\xrightarrow{\quad\psi_n\;\Longrightarrow \;\widetilde{\psi}_{n+1}\quad}&&\left(\;{\tiny\yng(2)}\;;\;{\tiny\yng(2,1,1)}\;\right)\qquad&&E_{2,4}+3\frac{\lambda}{N}-2\epsilon\\
&E_{4,2}-\frac{\lambda}{N}-\epsilon\qquad&&\left(\;{\tiny\yng(2,1,1)}\;;\;{\tiny\yng(1,1)}\;\right)&&\xrightarrow{\quad\psi_n\;\Longrightarrow \;\widetilde{\psi}_{n+1}\quad}&&\left(\;{\tiny\yng(1,1)}\;;\;{\tiny\yng(2,1,1)}\;\right)\qquad&&E_{2,4}+\frac{\lambda}{N}-2\epsilon\\
&E_{4,2}+\frac{\lambda}{N}-\epsilon\qquad&&\left(\;{\tiny\yng(2,2)}\;;\;{\tiny\yng(1,1)}\;\right)&&\xrightarrow{\quad\psi_n\;\Longrightarrow \;\widetilde{\psi}_{n+1}\quad}&&\left(\;{\tiny\yng(1,1)}\;;\;{\tiny\yng(2,2)}\;\right)\qquad&&E_{2,4}-\frac{\lambda}{N}\\
&E_{4,2}+\frac{\lambda}{N}+\epsilon\qquad&&\left(\;{\tiny\yng(3,1)}\;;\;{\tiny\yng(2)}\;\right)&&\xrightarrow{\quad\psi_n\;\Longrightarrow \;\widetilde{\psi}_{n+1}\quad}&&\left(\;{\tiny\yng(2)}\;;\;{\tiny\yng(3,1)}\;\right)\qquad&&E_{2,4}-\frac{\lambda}{N}+2\epsilon
\end{alignat*}
where $E_{a,b}=\frac{1}{2}\left|a-b\right|+\frac{\lambda}{2}\left(a-b\right)-\frac{1}{2N}\left(a-b\right)^2-\frac{\lambda}{2N^2}\left(a^2-b^2\right)+\frac{\frac{\lambda^2}{N^2}}{1+\frac{\lambda}{N}}\left(\frac{N}{2}-\frac{1}{2N}b^2\right)$

\section{Discussion and Open Issues}

We have given in this work a complete classification and nonlinear description of single and multi-trace operators in $W_N$ minimal CFT. In addition we have presented a (collective) Hamiltonian which generates the primary states at nonlinear level ( with $G=\sfrac{1}{N}$ as the coupling constant ). Consequently, our formulation can serve as a basis for the $\sfrac{1}{N}$ expansion of the model. It is the first step in direct re-construction of higher spin theory from large $N$ CFT.
This Hamiltonian shows analogies with matrix-vector type models is characterized by an extra dimension coming from the winding number. The theory exhibits locality, in terms of conjugate spacial $S_1$ coordinate. This is in accordance with the proposal originally due to Yin \cite{yin2012} that the complete duality in addition to higher-spins in AdS$_3$ space-time involves a further Kaluza-Klein type dimension with extra vector fields.

Our Hamiltonian can be expanded in various limits. In the t'Hooft limit $H_0$ and $H_1$ represent the  unpertururbed quadratic Hamiltonians. On the other hand, one can also consider the semiclassical limit  \cite{gaberdielgopakumar2012} and \cite{perlmutter2012}. In the semiclassical limit, the leading terms are $H_2+H_4+H_6$ up to global variables $K,\overline{K}$. These actually represent the pure matrix-vector model up to global variables. The other terms of order $\mathcal{O}\left(1\right)$ or lower such as $H_3+H_5$ play a role of perturbation in the semiclassical limit. It will be interesting to compare these interactions with recent study of correlation functions \cite{Hijano:2013fja}.

There are a number of important issues which were not taken into account in this basic construction. First of all, we have concentrated on the subspace of primary states of the $W_N$ model, and more restrictively the subset containing no derivatives. It is relatively simple to extend the construction to involve  primaries with derivatives and also all the descendants. For instance,
\begin{align}
\left(\;{\tiny\yng(2,1)}\;;\;0\;\right)\sim\psi_0^2\partial\bar{\partial} \psi_0-\psi_0\partial\psi_0\bar{\partial}\psi_0
\end{align}
Especially, a subsector $\left(\Lambda;0\right)$, which is closed under fusion product, is tractable. By acting $W_N$ generators on the primary, we can get descendants. This will be related to the creation operator of scalar field in the AdS$_3$ background.
\begin{align}
A_{n,m}\left(\sigma\right)\sim \oint dz\oint d\bar{z} L_{-1}^n\overline{L}_{-1}^m \mathcal{O}_{\Psi}\left(z,\bar{z};\sigma\right)
\end{align}
where $\mathcal{O}_{\Psi}\left(z,\bar{z};\sigma\right)$ is the vertex operator corresponding to $\Psi$. It is with this inclusion of derivatives and descendants that we see the full AdS$_3\times S_1$ space-time.

 We have also ignored the phenomenon of null-states. Their interpretation and role needs to be included. We can expect that the basic effective Hamiltonian that we have succeeded in constructing can point the way how it is to be done.  Various applications such as  to evaluation of  free energy~\cite{gaberdielgopakumarhartman2011,gaberdielcandu2012,perlmutter2012,Jain:2013py} and non-perturbative phenomena~\cite{Kraus2011,Ammon2011,Castro2012:1,Kraus2011:1,Banados2012,Kraus2012,Gaberdiel:2012yb,Castro2012,Banerjee:2012gh,Banerjee:2012aj} are obviously of high interest.
Finally , after our paper was posted there appeared a paper by Chang and Yin \cite{Chang:2013izp} which contains overlap with our work. We find agreement on the appearance of the extra dimension and the locality of the emergent theory.

\acknowledgments

This work evolved during the last year and a half benefiting from useful discussions with many collegues. We would like to thank Jean Avan, Sumit Das, Robert de Mello Koch, Sanjaye Ramgoolam, Soo-Jong Rey, Joao Rodrigues, Kewang Jin and  Qibin Ye for interest and useful comments. One of us (A.J) would like to thank Matthias Gaberdiel and Soo-Jong Rey for their hospitality at ETH, Zurich and SNU, respectively during some of the time that the work was done. This work was supported by the Department of Energy under contract DE-FG02-91ER40688.

\appendix
\section{Conformal Dimension}\label{conformaldimappendix}

The conformal dimension in \eqref{conformaldimension} can be expressed as
\begin{align}
\begin{split}
h\left(\Lambda_+;\Lambda_-\right)=\frac{1}{p}C\left(\Lambda_+\right)-\frac{1}{p+1}C\left(\Lambda_-\right)+\frac{1}{2}\left|\Lambda_+-\Lambda_-\right|^2
\end{split}\label{appendixeq1}
\end{align}

First of all, we can separate Casimir $C\left(\Lambda_\pm\right)$ into two parts of $R_\pm,S_\pm$ up to global variables, $r, s$. The first and the second terms of \eqref{appendixeq1} can be separated as
\begin{align}
\label{conformaldimensionreduced}
\begin{split}
C\left(\Lambda_\pm\right)=C\left(R_\pm\right)+C\left(S_\pm\right)+\frac{r_\pm s_\pm}{N}
\end{split}
\end{align}
where $r_\pm\equiv\left|R_\pm\right|,\; s_\pm\equiv\left|S_\pm\right|$.

\begin{figure}[tbp]
\centering
\includegraphics[width=13cm]{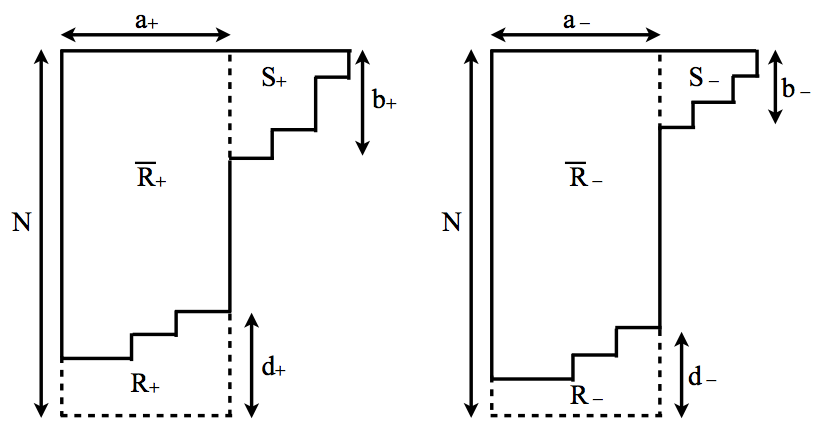}
\caption{\label{youngtableau2} Young tableau}
\end{figure}

Now, we can also separate the third term in the \eqref{appendixeq1} into two parts up to global variables.(See figure~\ref{youngtableau2})
\begin{align}
\begin{split}
&\frac{1}{2}\left|\Lambda_+-\Lambda_-\right|^2=\frac{1}{2}\left|S_+-S_-\right|^2+\frac{1}{2}\left|R_+-R_-\right|^2+\frac{1}{N}\left(s_+-s_-\right)\left(r_+-r_-\right)
\end{split}
\end{align}
Finally, we have
\begin{align}
\begin{split}
h\left(\Lambda_+;\Lambda_-\right)=&h\left(R_+;R_-\right)+h\left(S_+;S_-\right)+\frac{\lambda}{N^2}\left(r_+s_+-r_-s_-\right)+\frac{\frac{\lambda^2}{N^2}}{1+\frac{\lambda}{N}}\frac{r_-s_-}{N}\\
&+\frac{1}{N}\left(s_+-s_-\right)\left(r_+-r_-\right)
\end{split}
\end{align}

In summary , we can separate the conformal dimension of $\left(\Lambda_+;\Lambda_-\right)$ into conformal dimensions of $\left(R_+;R_-\right)$ and $\left(S_+;S_-\right)$ up to global variables. In detail, the conformal dimension is
\begin{align}
\begin{split}
h\left(\Lambda_+;\Lambda_-\right)=&\frac{\lambda}{2}\left(s_+-s_-+r_+-r_-\right)+\frac{1}{2}\sum_{i=1}^{N-1}\left[\left(s^+_{i}-s^-_{i}\right)^2+\left(r^+_{i}-r^-_{i}\right)^2\right]\\
&+\frac{\lambda}{2N}\left(D_{S_+}-D_{S_-}+D_{R_+}-D_{R_-}\right)-\frac{1}{2N}\left(s_+-s_--r_++r_-\right)^2\\
&-\frac{\lambda}{2N^2}\left(s_++s_--r_+-r_-\right)\left(s_+-s_--r_++r_-\right)\\
&+\frac{\frac{\lambda^2}{N^2}}{1+\frac{\lambda}{N}}\frac{N}{2}\left(s_-+r_-\right)+\frac{\frac{\lambda^2}{N^2}}{1+\frac{\lambda}{N}}\frac{1}{2}\left(D_{S_-}+D_{R_-}\right)-\frac{\frac{\lambda^2}{N^2}}{1+\frac{\lambda}{N}}\frac{1}{2N}\left(s_--r_-\right)^2
\end{split}
\end{align}
where
\begin{align*}
r^\pm_i\equiv&\left(\text{The number of boxes in the $i$th row of Young tableau }R_\pm\right)\\
s^\pm_i\equiv&\left(\text{The number of boxes in the $i$th row of Young tableau }S_\pm\right)
\end{align*}
Note that, when $B_\pm$ is used in equations, $B_\pm$ does not mean the total number of all boxes in $\Lambda_\pm$ which can be order $\mathcal{O}\left(N\right)$, but means the total number of boxes in $R_\pm$ and $S_\pm$ which are order $\mathcal{O}\left(1\right)$.
\begin{align}
B_\pm=\left|\Lambda_\pm\right|=\left|R_\pm\right|+\left|S_\pm\right|
\end{align}

\section{Three Point Function}\label{threepointfunctionappendix}
A primary field $\mathcal{O}_{\left(\Lambda_+;\Lambda_-\right)}$ of $W_N$ minimal model is normalized as

\begin{align}
\left<\mathcal{O}_{(\Lambda_+;\Lambda_-)}\mathcal{O}_{(\overline{\Lambda}_+;\overline{\Lambda}_-)}\right>=\frac{1}{\left|x_{12}\right|^{2\Delta}}
\end{align}
where $x_{ij}\equiv x_i-x_j$ and $\Delta=h\left(\Lambda_+;\Lambda_-\right)+\overline{h}\left(\Lambda_+;\Lambda_-\right)$.

Now, consider a three point function.
\begin{align}
\left<\mathcal{O}_{(\Lambda_+^1;\Lambda_-^1)}\mathcal{O}_{(\Lambda_+^2;\Lambda_-^2)}\mathcal{O}_{(\Lambda_+^3;\Lambda_-^3)}\right>=\frac{C_{3}\left({(\Lambda_+^1;\Lambda_-^1)},{(\Lambda_+^2;\Lambda_-^2)},{(\Lambda_+^3;\Lambda_-^3)}\right)}{\left|x_{12}\right|^{\Delta_1+\Delta_2-\Delta_3}\left|x_{23}\right|^{\Delta_2+\Delta_3-\Delta_1}\left|x_{31}\right|^{\Delta_3+\Delta_1-\Delta_2}}
\end{align}

\subsection{Examples of Three point functions}\label{threepointfunctionexample}

We calculated three point functions by following \cite{yin2011}. We can observe that the first order of three point function is the same as that of transposed Young tableaux. Accepting this transposition symmetry, we can get the first order of three point functions which are hard to calculate.
\begin{align*}
C_{3}&\left(\;\overline{{\tiny\yng(1)}}\;;\;\overline{{\tiny\yng(1)}}\;),(\;\overline{{\tiny\yng(1)}}\;;\;\overline{{\tiny\yng(1)}}\;),(\;{\tiny\yng(1,1)}\;;\;{\tiny\yng(1,1)}\;)\right)\\
&=1-\frac{\lambda^2}{2N^2}\left(\pi\cot\pi\lambda-\pi^2\lambda\csc^2\pi \lambda+2\gamma+2\psi(\lambda)+2\lambda\psi^{(1)}(\lambda)\right)+\mathcal{O}\left(\frac{1}{N^3}\right)\\
&\qquad\qquad\Updownarrow\qquad \mbox{Transpose}\\
C_{3}&\left((\;\overline{{\tiny\yng(1)}}\;;\;\overline{{\tiny\yng(1)}}\;),(\;\overline{{\tiny\yng(1)}}\;;\;\overline{{\tiny\yng(1)}}\;),(\;{\tiny\yng(2)}\;;\;{\tiny\yng(2)}\;)\right)\\
&=1+\frac{\lambda^2}{2N^2}\left(\pi\cot\pi\lambda-\pi^2\lambda\csc^2\pi \lambda+2\gamma+2\psi(\lambda)+2\lambda\psi^{(1)}(\lambda)\right)+\mathcal{O}\left(\frac{1}{N^3}\right)
\end{align*}
\begin{align*}
C_3&\left((\;\overline{{\tiny\yng(1)}}\; ;\; 0\;),(\;\overline{{\tiny\yng(1)}}\; ;\; \overline{{\tiny\yng(1)}} \;),(\;{\tiny\yng(1,1)}\; ;\;\tiny\yng(1)\;)\right)\\
&=\frac{1}{\sqrt{2}}-\frac{1}{2\sqrt{2}N}\left(2+\lambda\pi\cot\pi\lambda+2\lambda\gamma+2\lambda\psi(\lambda)\right)+\mathcal{O}\left(\frac{1}{N^3}\right)\\
&\qquad\qquad\Updownarrow\qquad \mbox{Transpose}\\
C_3&\left((\;\overline{{\tiny\yng(1)}}\; ;\; 0\;),(\;\overline{{\tiny\yng(1)}}\; ;\; \overline{{\tiny\yng(1)}} \;),(\;{\tiny\yng(2)}\; ;\;\tiny\yng(1)\;)\right)\\
&=\frac{1}{\sqrt{2}}+\frac{1}{2\sqrt{2}N}\left(2+\lambda\pi\cot\pi\lambda+2\lambda\gamma+2\lambda\psi(\lambda)\right)+\mathcal{O}\left(\frac{1}{N^3}\right)
\end{align*}
\begin{align*}
C_3&\left((\;\overline{{\tiny\yng(2)}}\; ;\; \overline{{\tiny\yng(1)}}\;),(\;\overline{{\tiny\yng(1)}}\; ;\; \overline{{\tiny\yng(1)}} \;),(\;{\tiny\yng(3)}\; ;\;{\tiny\yng(2)}\;)\right)\\
&=\sqrt{\frac{2}{3}}+\sqrt{\frac{2}{3}}\frac{1}{2N}\left(2+\pi\lambda\cot\pi\lambda+2\lambda\gamma+2\lambda\psi(\lambda)\right)+\mathcal{O}\left(\frac{1}{N^2}\right)
\end{align*}
\begin{align*}
C_3&\left((\;\overline{{\tiny\yng(2)}}; ;\; \overline{{\tiny\yng(1)}}\;),(\;\overline{{\tiny\yng(1)}}\; ;\; \overline{{\tiny\yng(1)}} \;),(\;{\tiny\yng(2,1)}\; ;\;{\tiny\yng(2)}\;)\right)\\
&=\frac{1}{2\sqrt{3}}-\frac{1}{2\sqrt{3}}\frac{2}{2N}\left(2+\pi\lambda\cot\pi\lambda+2\lambda\gamma+2\lambda\psi(\lambda)\right)+\mathcal{O}\left(\frac{1}{N^2}\right)\\
&\qquad\qquad\Updownarrow\qquad \mbox{Transpose}\\
C_3&\left((\;\overline{{\tiny\yng(1,1)}}; ;\; \overline{{\tiny\yng(1)}}\;),(\;\overline{{\tiny\yng(1)}}\; ;\; \overline{{\tiny\yng(1)}} \;),(\;{\tiny\yng(2,1)}\; ;\;{\tiny\yng(1,1)}\;)\right)\\
&=\frac{1}{2\sqrt{3}}-\frac{1}{2\sqrt{3}}\frac{1}{12N}\left(1+2\lambda\gamma+\pi\lambda\cot\pi \lambda+2\lambda\psi(\lambda)\right)+\mathcal{O}\left(\frac{1}{N^2}\right)
\end{align*}
\begin{align*}
C_3&\left((\;\overline{{\tiny\yng(1,1)}}; ;\; \overline{{\tiny\yng(1)}}\;),(\;\overline{{\tiny\yng(1)}}\; ;\; \overline{{\tiny\yng(1)}} \;),(\;{\tiny\yng(2,1)}\; ;\;{\tiny\yng(2)}\;)\right)\\
&=\frac{\sqrt{3}}{2}+\frac{\sqrt{3}}{2}\frac{\lambda^2}{2N^2}\left(\pi\cot\pi\lambda-\pi^2\lambda\csc^2\pi \lambda+2\gamma+2\psi(\lambda)+2\lambda\psi^{(1)}(\lambda)\right)+\mathcal{O}\left(\frac{1}{N^3}\right)\\
&\qquad\qquad\Updownarrow\qquad \mbox{Transpose}\\
C_3&\left((\;\overline{{\tiny\yng(2)}}; ;\; \overline{{\tiny\yng(1)}}\;),(\;\overline{{\tiny\yng(1)}}\; ;\; \overline{{\tiny\yng(1)}} \;),(\;{\tiny\yng(2,1)}\; ;\;{\tiny\yng(1,1)}\;)\right)\\
&=\frac{\sqrt{3}}{2}-\frac{\sqrt{3}}{2}\frac{\lambda^2}{2N^2}\left(\pi\cot\pi\lambda-\pi^2\lambda\csc^2\pi \lambda+2\gamma+2\psi(\lambda)+2\lambda\psi^{(1)}(\lambda)\right)+\mathcal{O}\left(\frac{1}{N^3}\right)
\end{align*}
\begin{align*}
C_3&\left((\;\overline{{\tiny\yng(2)}}; ;\; \overline{{\tiny\yng(1)}}\;),(\;\overline{{\tiny\yng(1)}}\; ;\; \overline{{\tiny\yng(1)}} \;),(\;{\tiny\yng(1,1,1)}\; ;\;{\tiny\yng(1,1)}\;)\right)=0
\end{align*}
\begin{align*}
C_3&\left((\;\overline{{\tiny\yng(1)}}\; ;\; 0\;),(\;\overline{{\tiny\yng(2)}}\; ;\; \overline{{\tiny\yng(2)}} \;),(\;{\tiny\yng(3)}\; ;\;{\tiny\yng(2)}\;)\right)=\frac{1}{\sqrt{3}}+\mathcal{O}\left(\frac{1}{N}\right)\\
&\qquad\qquad\Updownarrow\qquad \mbox{Transpose}\\
C_3&\left((\;\overline{{\tiny\yng(1)}}\; ;\; 0\;),(\;\overline{{\tiny\yng(1,1)}}\; ;\; \overline{{\tiny\yng(1,1)}} \;),(\;{\tiny\yng(1,1,1)}\; ;\;{\tiny\yng(1,1)}\;)\right)=\mathcal{O}\left(\frac{1}{N^4}\right)
\end{align*}
\begin{align*}
C_3&\left((\;\overline{{\tiny\yng(1)}}\; ;\; 0\;),(\;\overline{{\tiny\yng(2)}}\; ;\; \overline{{\tiny\yng(2)}} \;),(\;{\tiny\yng(2,1)}\; ;\;{\tiny\yng(1,1)}\;)\right)=0
\end{align*}
\begin{align*}
C_3&\left((\;\overline{{\tiny\yng(1)}}\; ;\; 0\;),(\;\overline{{\tiny\yng(2)}}\; ;\; \overline{{\tiny\yng(2)}} \;),(\;{\tiny\yng(2,1)}\; ;\;{\tiny\yng(2)}\;)\right)=\sqrt{\frac{2}{3}}+\mathcal{O}\left(\frac{1}{N}\right)\\
&\qquad\qquad\Updownarrow\qquad \mbox{Transpose}\\
C_3&\left((\;\overline{{\tiny\yng(1)}}\; ;\; 0\;),(\;\overline{{\tiny\yng(1,1)}}\; ;\; \overline{{\tiny\yng(1,1)}} \;),(\;{\tiny\yng(2,1)}\; ;\;{\tiny\yng(1,1)}\;)\right)=\sqrt{\frac{2}{3}}+\mathcal{O}\left(\frac{1}{N}\right)
\end{align*}
\begin{align*}
C_3&\left((\;\overline{{\tiny\yng(1)}}\; ;\; 0\;),(\;\overline{{\tiny\yng(2)}}\; ;\; \overline{{\tiny\yng(2)}} \;),(\;{\tiny\yng(1,1,1)}\; ;\;{\tiny\yng(1,1)}\;)\right)=0
\end{align*}

\section{Counting States}\label{appendixcountingstates}

In this section, we will count the number of states in $F_{m,\overline{m}}^{k,\overline{k}}$ and $Y_{s_+,s_-;r_+,r_-}$. Especially, $S_\pm$ is decoupled to $R_\pm$. Equivalently, $\psi_i,\omega_j$ are also decoupled to $\overline{\psi}_i, \overline{\omega}_j$. Therefore, it is sufficient to consider only $F_{m,0}^{k,0}$ and $Y_{s_+,s_-;0,0}$. Our claim is
\begin{align}
\left|F_{m,0}^{k,0}\right|=\left|Y_{m+k,k;0,0}\right|=\sum_{i=0}^k q\left(i,m\right)p\left(k-i\right)
\end{align}

\subsection{Partition of number}

Before starting the proof, define two functions. The number of partitions of non-negative integer $n$ is
\begin{align}
p\left(n\right)\equiv \left(\mbox{The number of partitions of } n\right)
\end{align}
For example, $p\left(0\right)=1,\; p\left(1\right)=1,\;p\left(2\right)=2,\; p\left(3\right)=3,\;p\left(4\right)=5\cdots$. In addition, we can restrict the number of integers that form partition of an integer. Consider all partitions of a non-negative integer $n$ which have at most $m$ positive integers as partition elements.
\begin{align}
\begin{split}
q\left(n,m\right)&\equiv\left|\left\{(x_1,\cdots,x_m)\left|\sum_{i=1}^m x_i=n,\quad x_1\geqq x_{2}\geqq \cdots\geqq x_m\geqq 0,\quad x_i\in \mathbb{Z}\right.\right\}\right|\\
&=\left(\mbox{The number of partitions of $n$ by $m$ non-negative integers}\right)
\end{split}
\end{align}
For example, 
\begin{align*}
&q\left(3,2\right)=2,\qquad q\left(3,1\right)=1,\qquad q\left(4,3\right)=4,\qquad q\left(4,2\right)=3\\
&q\left(4,1\right)=1,\qquad q\left(5,4\right)=6,\qquad q\left(5,3\right)=5,\qquad q\left(5,2\right)=3
\end{align*}
Especially, $q\left(n,m\right)$ is related to $p\left(n\right)$ through
\begin{align}
q\left(n,n\right)=p\left(n\right)\qquad \left(\;n>0\;\right)
\end{align}

\subsection{$F_{m,0}^{k,0}$}

\begin{align}
\mbox{For} \quad \prod_{i=1}^m\psi_{a_i}\prod_j W_{b_j}\in F_{m,0}^{k,0}\;,\qquad \sum_j b_j=k-\sum_{i=1}^m a_i
\end{align}
For each $\sum_{i=1}^m a_i=0,1,\cdots,k$,
we can get
\begin{align}
\left|F_{m,0}^{k,0}\right|=\sum_{i=0}^k q\left(i,m\right)p\left(k-i\right)
\end{align}

\subsection{$Y_{k+m,k;0,0}$}

This proof is complicated. Thus, we will divide it into two parts, modifying the problem and checking bijection correspondence.

\subsubsection*{Modification}

\begin{align}
\left(\Lambda_+;\Lambda_-\right)=\left(S_+;S_-\right)\quad,\qquad\mbox{where } S_+=k+m, S_-=k
\end{align}

In the same way as before, we will count the number of states by counting ways to add $m$ boxes (at most one box in each row) to all possible Young tableaux $S_-$ with $k$ boxes. 

\begin{align}
{\yng(4,2,2,1,1,0)}\qquad+\qquad{\yng(1,1,0,1,0,1)}\qquad\longrightarrow \qquad{\yng(5,3,2,2,1,1)}
\end{align}

For example, add 4 boxes to Young tableau with 10 boxes. First, we can choose one Young tableau with 10 boxes. Then, we can choose a way to add 4 boxes to the chosen Young tableau. We can arrange these additional 4 boxes in one column like the above figure. 

Alternatively, we can count the same situation in a different way. We can choose an array of additional 4 boxes first. Then, we can choose suitable Young tableau with 10 boxes. We can not choose arbitrary Young tableaux. For instance,
\begin{align}
\label{examplearray}
{\yng(4,2,2,1,1,0)}\quad + \quad{\yng(1,1,0,1,0,1)}\quad :\mbox{Possible},\qquad
{\yng(4,2,2,2,1,0)}\quad + \quad{\yng(1,1,0,1,0,1)}\quad :\mbox{Impossible}
\end{align}
For fixed addition boxes, the first Young tableau is possible but we cannot choose the second Young tableau. Thus, through this example, we can guess relation between the structure of additional array and the possible Young tableaux.

\begin{figure}[tbp]
\centering
\includegraphics[width=6cm]{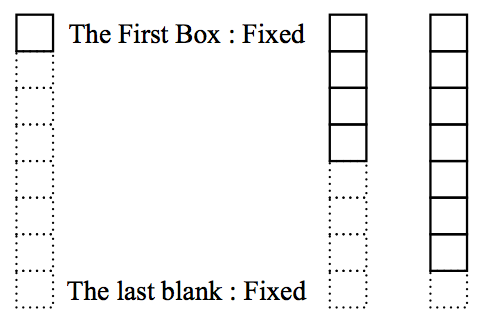}
\caption{\label{cluster} Clusters}
\end{figure}

In order to analyze this relation, define cluster. A cluster is a vertical array of boxes and blanks. Every cluster starts with a box and ends with a blank. And cluster can be filled with boxes from the first box. Hence, the minimum length of a cluster is 2.\footnote{The first cluster is an exception. The minimum length of the first cluster is 1 because it can start with a blank.} In figure~\ref{cluster}, we can see three examples of clusters with length 8.

Any additional array of boxes and blanks can be expressed as a sequence of clusters under the condition that the first cluster does not have the first fixed box (That is, the first cluster can start with a blank.). Especially, the last cluster can be considered to have an infinite series of blanks. Therefore, we will ignore the last cluster from now on. 

\begin{figure}[tbp]
\centering
\includegraphics[width=8cm]{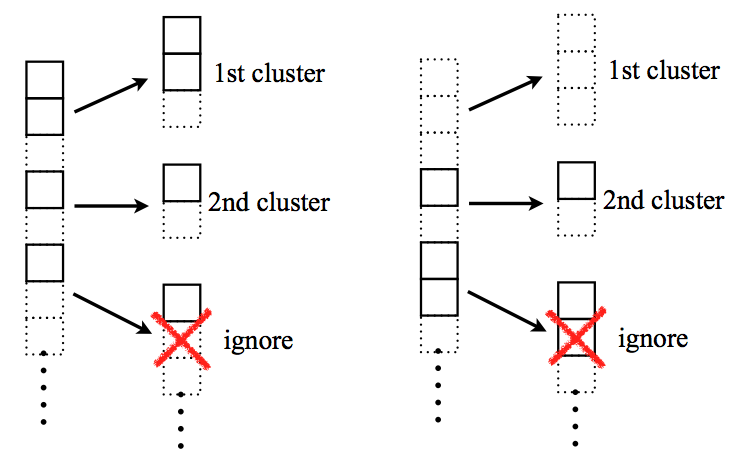}
\caption{\label{clusterexample} Examples for expression of arrays of boxes and blanks with clusters}
\end{figure}

For example, in figure~\ref{clusterexample}, the first one corresponds to the previous example~\eqref{examplearray} of array with four boxes. Both of two examples are consist of three clusters including the last cluster. But, we will ignore the last clusters so that we will consider only the first two clusters of them, respectively. Especially, the second example shows that the first cluster does not have the starting box. 

This decomposition of an array into cluster provides good information about all possible Young tableaux for the given array. If a cluster starts at $i$th row, we can only choose Young tableaux with a corner at $(i-1)$th row. On the other hands, since the first cluster always starts from the first row, it does not impose any restriction on Young tableaux. In the above first example, suitable Young tableaux should have corners at the 3rd and the 5th row.

Now, when we decompose array of boxes and blanks into a sequence of clusters, let the starting position of $i$th cluster be $a_i$. Then, candidate Young tableaux must have corners at $(a_i-1)$th row. Thus, the number of possible Young tableaux with $n$ boxes is
\begin{align}
p\left(n-\sum_{i}(a_i-1)\right)
\end{align}

Therefore, we need to count the number of configurations of arrays with $l=\sum_{i}(a_i-1)$ fixed. In the next section, it will be shown that this number is $q\left(l,m\right)$ by considering a bijection map between the configurations of array and restricted partitions of $l$.$\left(l=0,1,2,\cdots,k\right)$

\subsubsection*{Bijection map}

First of all, let's define a map from an array of boxes and blanks to a partition of a number. Suppose an array is decomposed into $(n+1)$ clusters. In addition, let $\sum_i (a_i-1)=l$ where the $i$th cluster starts at the $a_i$th row.

\begin{figure}[tbp]
\centering
\includegraphics[width=3.5cm]{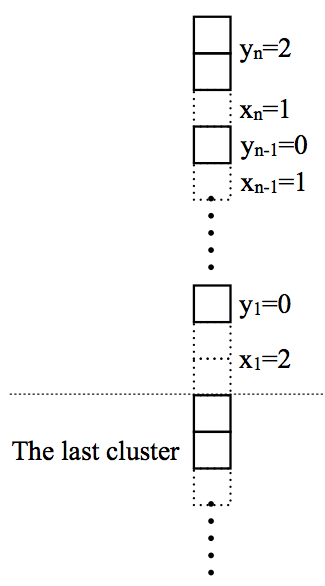}
\caption{\label{exampleofxy} Examples for $x_i, y_i$}
\end{figure}

Define non-negative integers, $x_i, y_i$ ($i=1,\cdots,n$) such that
\begin{alignat*}{3}
&x_i&&=\left(\mbox{The number of blanks in the $n-i+1$th cluster}\right)\qquad &&(i=1,\cdots,n)\\
&y_i&&=\left(\mbox{The number of boxes in the $n-i+1$th cluster}\right)-1\qquad &&(i=1,\cdots,n-1)\\
&y_n&&=\left(\mbox{The number of boxes in the 1st cluster}\right)&&
\end{alignat*}
We will ignore $y_{0}$ and $x_{0}=\infty$ in the last cluster. For example, see the figure~\ref{exampleofxy}.

Then, this configuration of the array can be mapped to a partition of $l$ by the following way.
\begin{align}
\begin{split}
l=&\underbracket{\left(x_1+x_2+x_3+x_4+\cdots+x_n\right)}_{\mbox{one number}}+\underbracket{1+1+\cdots+1}_{y_1}\\
&+\underbracket{\left(1+x_2+x_3+x_4+\cdots+x_n\right)}_{\mbox{one number}}+\underbracket{2+2+\cdots+2}_{y_2}\\
&+\underbracket{\left(2+x_3+x_4+\cdots+x_n\right)}_{\mbox{one number}}+\underbracket{3+3+\cdots+3}_{y_3}\\
&\qquad\qquad\vdots\qquad\qquad\qquad\qquad \vdots\\
&+\underbracket{\left(n-1+x_n\right)}_{\mbox{one number}}+\underbracket{n+n+\cdots+n}_{y_n}
\end{split}
\end{align}
In order to confirm that this map gives a partition of $l$, we can calculate the right hand side, 
\begin{align}
\mbox{RHS}=\sum_{i=1}^n \left[i\left(x_i+y_i+1\right)\right]-n=\sum_{i=1}^n\left[\left\{\sum_{j=n-i+1}^{n}\left(x_j+y_j+1\right)\right\}-1\right]
\end{align}
Especially, $\sum_{j=n-i+1}^n\left(x_j+y_j+1\right)$ is the position of the first box of the $i$th cluster, $a_i$. Thus, since we selected arrays with $\sum_i (a_i-1)=l$, the right hand side is indeed a partition of $l$.

Before getting inverse map, let's see a property of this partition. Since $x_i$ are positive integers and $y_i$ are non-negative integers, we can arrange them in non-increasing order.
\begin{align}
\begin{split}
&\underbracket{\left(x_1+\cdots+x_n\right)}_{\mbox{one number}}\geqq \underbracket{\left(1+x_2+\cdots+x_n\right)}_{\mbox{one number}}\geqq\cdots\geqq \underbracket{\left(n-1+x_n\right)}_{\mbox{one number}}\geqq \underbracket{n\geqq n\geqq\cdots\geqq n}_{y_n}\geqq\cdots\\
&\cdots\geqq\underbracket{2\geqq\cdots\geqq 2}_{y_2}\geqq\underbracket{1\geqq\cdots\geqq 1}_{y_1}
\end{split}\label{series of partition}
\end{align}
Moreover, note that the $i$th number in the series \eqref{series of partition} is greater than $i-1$ only for $i=1,2,\cdots,n$. For instance,
\begin{align}
\begin{split}
\underbracket{\left(x_1+\cdots+x_n\right)}_{\mbox{one number}}\quad>&\quad 0\\
\underbracket{\left(1+x_2+\cdots+x_n\right)}_{\mbox{one number}}\quad>&\quad1\\
\vdots\;&\\
n\quad>&\quad n-1
\end{split}
\end{align}
On the other hands, the $j$th numbers ($j=n+1,n+2,\cdots$) do not satisfy this condition.
\begin{align}
n\ngtr n, \qquad\mbox{etc}
\end{align}

Now, consider how to invert this map. For given partition of $l$, we will construct sequence of clusters. Suppose we have partition of $l$. By arranging it in non-increasing order,
\begin{align}
l=z_1+z_2+\cdots +z_p\qquad \mbox{where}\quad z_1\geqq z_2\geqq\cdots \geqq z_p>0
\end{align}
for some positive integer $p$. Compare this ordered partition with increasing sequence $0,1,2,\cdots$. That is, compare $z_i$ and $i-1$. There exist minimum integer $n$ such that 
\begin{align}
z_n> n-1 \qquad\mbox{and}\qquad z_{n+1}\ngtr(n+1)-1=n
\end{align}
Then, we can set
\begin{align}
\begin{split}
z_1&=\left(x_1+\cdots+x_n\right)\\
z_2&=\left(1+x_2+\cdots+x_n\right)\\
\vdots&\qquad\qquad\vdots\\
z_n&=\left(n-1+x_n\right)
\end{split}
\end{align}
and
\begin{align}
\left\{z_{n+1},z_{n+2},\cdots,z_p\right\}=\left\{\underbracket{n,n,\cdots,n}_{y_n},\cdots,\underbracket{2,\cdots,2}_{y_2},\underbracket{1,\cdots,1}_{y_1}\right\}
\end{align}
By this procedure, we can recover all positive integers, $x_i$ and non-negative integers, $y_i$, which corresponds to the original array. Therefore, this is the inverse map.

The map from arrays to restricted partitions and its inverse map are well-defined. Therefore, we can conclude that the number of possible configuration of array with $\sum_i (a_i-1)=l$ fixed is same as the number of partition $l$. (where the $i$th cluster starts at the $a_i$th row)

Before finishing proof, we have to check a very important property. Consider the number of positive integers in the partition.
\begin{align}
\left(\mbox{The number of positive integers in the partition}\right)=n+\sum_{i=1}^n y_i=y_n+\sum_{i=1}^{n-1}y_i+1
\end{align}

\begin{figure}[tbp]
\centering
\includegraphics[width=9cm]{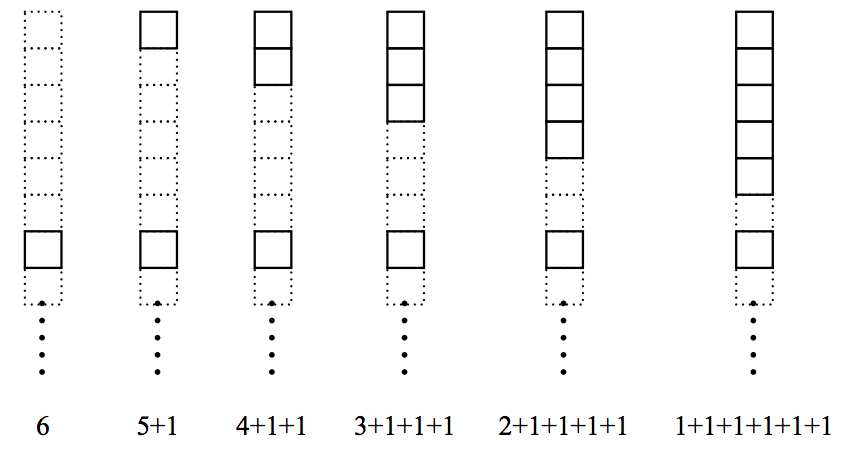}
\includegraphics[width=6cm]{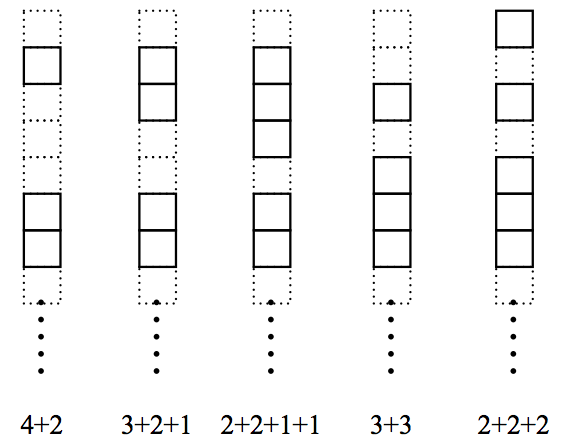}
\caption{\label{exampleof6} Examples of mapping : $l=6$}
\end{figure}

Recall that the last cluster has at least one box and $y_n$ is the number of boxes in the 1st cluster while $y_i+1$ is the number of boxes in the $n-i+1$th cluster $\left(i=1,2,\cdots,n-1\right)$. Hence, $\left(n+\sum_{i=1}^n y_i\right)$ is the minimum number of boxes for the corresponding configuration of array. In other words, the number of positive integers in the partition should be  less than or equal to the number of all boxes in the array. For example, there are 11 partitions of 6 (figure~\ref{exampleof6}). 6 itself is a partition of 6 and this partition has one positive integer. Thus, corresponding array should have at least one box. In addition, $5+1$ and $4+2$ have two positive integers, respectively. And, corresponding array should have at least two boxes. Moreover, arrays corresponding to $4+1+1$, $3+2+1$ and $2+2+2$ must have at least 3 boxes.

Therefore, a set of arrays (with $\sum_i (a_i-1)=l$ fixed and total number of boxes in the array, $m$ fixed) is bijectively mapped to a set of partitions of $l$ by  $m$ non-negative integers. Thus, the number of elements in this set of arrays is $q\left(l,m\right)$.

Summarizing these results,
\begin{align}
\left|Y_{m+k,k;0,0}\right|=\sum_{i=0}^k q\left(i,m\right)p\left(k-i\right)
\end{align}

% The bibliography will probably be heavily edited during typesetting.
% We'll parse it and, using the arxiv number or the journal data, will
% query inspire, trying to verify the data (this will probalby spot
% eventual typos) and retrive the document DOI and eventual errata.
% We however suggest to always provide author, title and journal data:
% in short all the informations that clearly identify a document.


\begin{thebibliography}{99}

\bibitem{vasiliev1987:1}
E.S.~Fradkin and M.A.~Vasiliev, \emph{Cubic interaction in extended theories of massless higher spin fields}, \emph{Phys. Phys. B} {\bf 291} (1987) 141.

\bibitem{vasiliev1987:2}
E.S.~Fradkin and M.A.~Vasiliev, \emph{Candidate to the Role of Higher Spin Symmetry}, \emph{Annals Phys.} {\bf 177} (1987) 63.

\bibitem{vasiliev1987:3}
E.S.~Fradkin and M.A.~Vasiliev, \emph{On the gravitational interaction of massless higher spin fields}, \emph{Phys. Lett. B} {\bf 189} (1987) 89. 

\bibitem{vasiliev1999}
M.A.~Vasiliev, \emph{Higher Spin Gauge Theories: Star-Product and AdS Space}, in M.A. Shifman (ed.), 'The many faces of the superworld' [arXiv:hep-th/9910096].

\bibitem{vasiliev2003}
M.A.~Vasiliev, \emph{Nonlinear Equations for Symmetric Massless Higher Spin Fields in (A)dS(d)}, \emph{Phys. Lett. B} {\bf 567} (2003) 139 [arXiv:hep-th/0304049].

\bibitem{klebanov2002}
I.R.~Klebanov and A.M.~Polyakov, \emph{AdS Dual of the Critical O(N) Vector Model}, \emph{Phys. Lett. B} {\bf 550} (2002) 213 [arXiv:hep-th/0210114].

\bibitem{sezgin2002}
E.~Sezgin and P.~Sundell, \emph{Massless Higher Spins and Holography}, \emph{Nucl. Phys. B} {\bf 644} (2002) 303 [Erratum-ibid. B {\bf 660} (2003) 403] [arXiv:hep-th/0205131].

\bibitem{giombi2010}
S.~Giombi and X.~Yin, \emph{Higher Spin Gauge Theory and Holography: The Three-Point Functions}, \emph{JHEP} {\bf 1009} (2010) 115 [arXiv:0912.3462~[hep-th]].

\bibitem{giombi2011:1}
S.~Giombi and X.~Yin, \emph{Higher Spins in AdS and Twistorial Holography}, \emph{JHEP} {\bf 1104} (2011) 086 [arXiv:1004.3736~[hep-th]].

\bibitem{Das:2003vw} 
S.~R.~Das and A.~Jevicki, \emph{Large $N$ Collective Fields and Holography,} \emph{Phys. Rev. D} {\bf 68} (2003) 044011 [arXiv:hep-th/0304093].

\bibitem{antal2010}
R.d.M.~Koch, A.~Jevicki, K.~Jin and J.P.~Rodrigues, \emph{AdS$_4$/CFT$_3$ Construction from Collective Fields}, \emph{Phys. Rev. D} {\bf 83} (2010) 025006 [arXiv:1008.0633~[hep-th]].

\bibitem{giombi2011:2}
S.~Giombi and X.~Yin, \emph{On Higher Spin Gauge Theory and The Critical $O(N)$ Model}, (2011) 096 [arXiv:1105.4011~[hep-th]].

\bibitem{antal2011}
A.~Jevicki, K.~Jin and Q.~Ye, \emph{Collective Dipole Model of AdS/CFT and Higher Spin Gravity}, \emph{J. Phys. A} {\bf 44} (2011) 465402 [arXiv:1106.3983~[hep-th]].


\bibitem{Giombi:2012ms} 
S.~Giombi and X.~Yin, \emph{The Higher Spin/Vector Model Duality,} (2012) [arXiv:1208.4036~[hep-th]].

\bibitem{Maldacena2011}
J.~Maldacena and A.~Zhiboedov, \emph{Constraining Conformal Field Theories with A Higher Spin Symmetry,} (2011) [arXiv:1112.1016~[hep-th]].

\bibitem{Gelfond:2013xt} 
O.~A.~Gelfond and M.~A.~Vasiliev, \emph{Operator algebra of free conformal currents via twistors,} (2013) [arXiv:1301.3123~[hep-th]].




 



\bibitem{henneaux2010}
M.~Henneaux and S.J.~Rey, \emph{Nonlinear W$_\infty$ Algebra as Asymptotic Symmetry of Three-Dimensional Higher Spin AdS Gravity}, \emph{JHEP} {\bf 1012} (2010) 007 [arXiv:1008.4579~[hep-th]].

\bibitem{campoleoni2010}
A.~Campoleoni, S.~Fredenhagen, S.~Pfenninger and S.~Theisen, \emph{Asymptotic symmetries of three-dimensional gravity coupled to higher-spin fields}, \emph{JHEP} {\bf 1011} (2010) 007 [arXiv:1008.4744~[hep-th]].

\bibitem{campoleoni2011}
A.~Campoleoni, S.~Fredenhagen and S.~Pfenninger, \emph{Asymptotic W-symmetries in three-dimensional higher-spin gauge theories}, \emph{JHEP} {\bf 1109} (2011) 113 [arXiv:1107.0290~[hep-th]].

\bibitem{gaberdielhartman2011}
M.~R.~Gaberdiel and T.~Hartman, \emph{Symmetries of Holographic Minimal Models}, \emph{JHEP} {\bf 1105} (2011) 031 [arXiv:1101.2910~[hep-th]].

\bibitem{gaberdielsaha2011}
M.~R.~Gaberdiel, R.~Gopakumar and A.~Saha, \emph{Quantum W-symmetry in AdS$_3$}, \emph{JHEP} {\bf 1102} (2011) 004 [arXiv:1009.6087~[hep-th]].

\bibitem{gaberdielgopakumar2011}
M.~R.~Gaberdiel and R.~Gopakumar, \emph{An AdS$_3$ Dual for Minimal Model CFTs}, \emph{Phys. Rev. D} {\bf 83} (2011) 066007 [arXiv:1011.2986~[hep-th]].

\bibitem{gaberdielgopakumarhartman2011}
M.~R.~Gaberdiel, R.~Gopakumar, T.~Hartman and S.~Raju, \emph{Partition Functions of Holographic Minimal Models}, \emph{JHEP} {\bf 1108} (2011) 077 [arXiv:1106.1897~[hep-th]].

\bibitem{yin2011:1}
C.~M.~Chang and X.~Yin, \emph{Higher Spin Gravity with Matter in AdS3 and Its CFT Dual,} (2011) [arXiv:1106.2580~[hep-th]].


\bibitem{gaberdielcandu2012}
C.~Candu and M.~R.~Gaberdiel, \emph{Supersymmetric holography on AdS$_3$}, (2012) [arXiv:1203.1939~[hep-th]].

\bibitem{Gaberdiel:2012uj} 
  M.~R.~Gaberdiel and R.~Gopakumar, \emph{Minimal Model Holography,} \emph{J. Phys. A} {\bf 46} (2013) 214002 [arXiv:1207.6697~[hep-th]].

\bibitem{Corley:2001zk}
S.~Corley, A.~Jevicki and S.~Ramgoolam, \emph{Exact Correlators of Giant Gravitons from dual N=4 SYM theory,} \emph{Adv. Theor. Math. Phys.} {\bf 5} (2002) 809 [arXiv:hep-th/0111222].

\bibitem{Lin:2004nb} 
H.~Lin, O.~Lunin and J.~M.~Maldacena, \emph{Bubbling AdS space and 1/2 BPS geometries,} \emph{JHEP} {\bf 0410} (2004) 025 [arXiv:hep-th/0409174].

\bibitem{Jevicki:1998rr} 
  A.~Jevicki and S.~Ramgoolam, \emph{Noncommutative gravity from the AdS / CFT correspondence,} \emph{JHEP} {\bf 9904} (1999) 032 [arXiv:hep-th/9902059].

\bibitem{bouwknegt1993}
P.~Bouwknegt and K.~Schoutens, \emph{$W$-symmetry in Conformal Field Theory}, \emph{Phys. Rept.} {\bf 223} (1993) 183 [arXiv:hep-th/9210010].

\bibitem{raju2011}
K.~Papadodimas and S.~Raju, \emph{Correlation Functions in Holographic Minimal Models}, \emph{Nucl. Phys. B} {\bf 856} (2011) 607 [arXiv:1108.3077~[hep-th]].

\bibitem{yin2011}
C.~M.~Chang and X.~Yin, \emph{Correlators in $W_N$ Minimal Model Revisited,} (2011) [arXiv:1112.5459~[hep-th]].

\bibitem{yin2012}
X.~Yin, \emph{Dissecting holography with higher spins, in Proceedings of the Ginzburg Conference on Physics}, Lebedev Institute, Moscow, May 2012.

\bibitem{antal1992}
A.~Jevicki, \emph{Non-perturbative collective field theory}, \emph{Nucl. Phys. B} {\bf 376} (1992) 75.

\bibitem{antal1996}
A.~Jevicki and J.~Avan, \emph{Collective field theory of the matrix-vector models}, \emph{Nucl. Phys. B} {\bf 469} (1996) 469 [arXiv:hep-th/9512147v2].


\bibitem{Avan:1995sp} 
J.~Avan and A.~Jevicki, \emph{Collective field theory of the matrix vector models}, \emph{Nucl. Phys. B} {\bf 469} (1996) 287 [hep-th/9512147].



\bibitem{Avan:1996vi} 
J.~Avan, A.~Jevicki and J.~Lee, \emph{Field theory of SU(R) spin Calogero-Moser models}, \emph{Nucl. Phys. B} {\bf 486} (1997) 650 [arXiv:hep-th/9607083].



\bibitem{antal2013}
A.~Jevicki and J.~Yoon, \emph{Field Theory of Characters,} \emph{Brown rept. HET-1638} (2013) (in preparation).


\bibitem{Das:1990kaa} 
S.~R.~Das and A.~Jevicki, \emph{String Field Theory And Physical Interpretation Of D = 1 Strings}, \emph{Mod. Phys. Lett. A} {\bf 5} (1990) 1639.


\bibitem{gaberdielgopakumar2012}
M.~R.~Gaberdiel and R.~Gopakumar, \emph{Triality in Minimal Model Holography}, (2012) [arXiv:1205.2472~[hep-th]].

\bibitem{perlmutter2012}
E.~Perlmutter, T.~Prochazka and J.~Raeymaekers, \emph{The semiclassical limit of W$_N$ CFTs and Vasiliev theory}, (2012) [arXiv:1210.8452~[hep-th]].


\bibitem{Hijano:2013fja} 
  E.~Hijano, P.~Kraus and E.~Perlmutter, \emph{Matching four-point functions in higher spin AdS$_3$/CFT$_2$}, arXiv:1302.6113~[hep-th]



\bibitem{Jain:2013py} 
S.~Jain, S.~Minwalla, T.~Sharma, T.~Takimi, S.~R.~Wadia and S.~Yokoyama, \emph{Phases of large $N$ vector Chern-Simons theories on $S^2 \times S^1$,} (2013) [arXiv:1301.6169~[hep-th]].



\bibitem{Kraus2011}
M.~Gutperle and P.~Kraus, \emph{Higher Spin Black Holes,} \emph{JHEP} {\bf 1105} (2011) 022 [arXiv:1103.4304~[hep-th]].

\bibitem{Ammon2011}
M.~Ammon, M.~Gutperle, P.~Kraus and E.~Permutter, \emph{Spacetime Geometry in Higher Spin Gravity,} \emph{JHEP} {\bf 1110} (2011) 053 [arXiv:1106.4788~[hep-th]].

\bibitem{Castro2012:1}
A.~Castro, E.~Hijano, A.~Lepage-Jutier and A.~Maloney, \emph{Black Holes and Singularity Resolution in Higher Spin Gravity,} \emph{JHEP} {\bf 1201} (2012) 031 [arXiv:1110.4117~[hep-th]].

\bibitem{Kraus2011:1}
P.~Kraus and E.~Perlmutter, \emph{Partition functions of higher spin black holes and their CFT duals,} \emph{JHEP} {\bf 1111} (2011) 061 [arXiv:1108.2567 [hep-th]].


\bibitem{Gaberdiel:2012yb} 
M.~R.~Gaberdiel, T.~Hartman and K.~Jin, \emph{Higher Spin Black Holes from CFT,} \emph{JHEP} {\bf 1204} (2012) 103 [arXiv:1203.0015 [hep-th]].

\bibitem{Banados2012}
M.~Banados, R.~Canto and S.~Theisen, \emph{The Action for higher spin black holes in three dimensions,} (2012) [arXiv:1204.5105~[hep-th]].

\bibitem{Kraus2012}
P.~Kraus and E.~Perlmutter, \emph{Probing higher spin black holes,} (2012) [arXiv:1209.4937~[hep-th]].

\bibitem{Castro2012}
A.~Castro, R.~Gopakumar, M.~Gutperle and J.~Raeymaekers, \emph{Conical Defects in Higher Spin Theories}, \emph{JHEP} {\bf 1202} (2012) 096 [arXiv:1111.3381~[hep-th]].

\bibitem{Banerjee:2012gh} 
S.~Banerjee, S.~Hellerman, J.~Maltz and S.~H.~Shenker, \emph{Light States in Chern-Simons Theory Coupled to Fundamental Matter,} (2012) [arXiv:1207.4195~[hep-th]].

\bibitem{Banerjee:2012aj} 
  S.~Banerjee, A.~Castro, S.~Hellerman, E.~Hijano, A.~Lepage-Jutier, A.~Maloney and S.~Shenker,
  \emph{Smoothed Transitions in Higher Spin AdS Gravity,} [arXiv:1209.5396~[hep-th]].


\bibitem{Chang:2013izp} 
  C.~-M.~Chang and X.~Yin, \emph{A semi-local holographic minimal model}, arXiv:1302.4420~[hep-th].
  
  
  \bibitem{Hijano:2013fja} 
  E.~Hijano, P.~Kraus and E.~Perlmutter, \emph{Matching four-point functions in higher spin AdS$_3$/CFT$_2$}, arXiv:1302.6113~[hep-th].
  
  
  

% Please avoid comments such as "For a review'', "For some examples",
% "and references therein" or move them in the text. In general,
% please leave only references in the bibliography and move all
% accessory text in footnotes.

% Also, please have only one work for each \bibitem.


\end{thebibliography}
\end{document}